\documentclass[twocolumn,10pt]{IEEEtran}
\usepackage{amsmath}
\usepackage{amssymb,amsthm}
\usepackage{graphicx}
\usepackage{psfrag}
\usepackage{url}
\usepackage{bm}
\usepackage{tikz,subfigure}

\urldef\rmmail\url{mori@is.titech.ac.jp}
\urldef\ttmail\url{tt@i.kyoto-u.ac.jp}

\title{Source and Channel Polarization over Finite Fields and Reed-Solomon Matrices}
\author{Ryuhei~Mori~\IEEEmembership{Member,~IEEE} and Toshiyuki Tanaka~\IEEEmembership{Member,~IEEE}%
\thanks{This paper was presented in part at~\cite{mori2010cpq,mori2010non}.
The work of R. Mori was supported by the Grant-in-Aid for Scientific Research for JSPS Fellows (22$\cdot$5936).
The work of T. Tanaka was supported by the Grant-in-Aid for Scientific Research (C), JSPS, Japan (22560375).}%
\thanks{R. Mori is with the Department of Mathematical and Computing Science, Graduate School of Information Science and Engineering,
Tokyo Institute of Technology, Shibaura, Minato-ku, Tokyo, 108-0023 Japan
(e-mail: \rmmail).}%
\thanks{T. Tanaka is with the Department of Systems Science, Graduate School of Informatics,
Kyoto University, Yoshida Hon-machi, Sakyo-ku, Kyoto-shi, Kyoto, 606-8501 Japan
(e-mail: \ttmail).}%
}

\allowdisplaybreaks[2]

\newtheorem{theorem}{Theorem}
\newtheorem{lemma}[theorem]{Lemma}
\newtheorem{corollary}[theorem]{Corollary}
\newtheorem{proposition}[theorem]{Proposition}
\theoremstyle{definition}
\newtheorem{definition}[theorem]{Definition}
\theoremstyle{remark}
\newtheorem{remark}{Remark}
\def\sasoglu{{{\c S}a{\c s}o{\u g}lu}}
\def\statq{invariant under any permutation of symbols in the a posteriori distribution}

\begin{document}
\maketitle

\begin{abstract}
Polarization phenomenon over any finite field $\mathbb{F}_{q}$ with size $q$ 
being a power of a prime is considered.
This problem is a generalization of the original proposal 
of channel polarization by Ar{\i}kan for the binary field, 
as well as its extension to a prime field by \sasoglu, Telatar, and Ar{\i}kan. 
In this paper, a necessary and sufficient condition of a matrix over a finite field $\mathbb{F}_q$ is shown under
which any source and channel are polarized.
Furthermore, the result of the speed of polarization for the binary alphabet obtained by Ar{\i}kan and Telatar is generalized to arbitrary finite field.
It is also shown that the asymptotic error probability of polar codes is improved by using the Reed-Solomon matrices,
which can be regarded as a natural generalization of the $2\times 2$ binary matrix used in the original proposal by Ar{\i}kan.
\end{abstract}

\section{Introduction}
\IEEEPARstart{A}{r{\i}kan} introduced the method of source and channel polarization which gives
efficient capacity-achieving binary source and channel codes,
respectively~\cite{5075875}.
\sasoglu~et~al.\ generalized the polarization phenomenon to non-binary alphabets whose size is a prime~\cite{sasoglu2009pad}.
They showed an example of a quaternary channel which is not polarized by Ar{\i}kan's $2\times2$ matrix.
Although there are channels not polarized by Ar{\i}kan's $2\times 2$ matrix for non-prime alphabets, 
one can argue that any channel is polarized in a weaker sense, 
as discussed in~\cite{abbe2012polar}.
From this observation, the symmetric capacity of any non-binary channel is efficiently achievable by directly using the channel polarization phenomenon~\cite{sasoglu2009pad}, \cite{abbe2012polar},
\cite{park2013polar}, \cite{sahebi2011multilevel}.
In~\cite{sasoglu2012polar}, a sufficient condition for a matrix over a ring $\mathbb{Z}/q\mathbb{Z}$ is shown on which any $q$-ary channel is polarized.
In this paper, we study the polarization phenomenon caused by matrices over finite fields.

The contributions of this paper are threefold.  
The first contribution is that 
we give a complete characterization as to 
whether an $\ell\times\ell$ matrix over a finite field 
gives rise to polarization.  
This extends the result on the binary field 
by Korada et al.~\cite{korada2010polar} to a general finite field.  
The second contribution is that we characterize the asymptotic speed of 
polarization in terms of the matrix used.  
This is again an extension of the result on the binary field 
by Korada et al.~\cite{korada2010polar} to a general finite field.  
The third contribution of this paper is that 
we provide an explicit construction of an $\ell\times\ell$ matrix, 
which is based on  the Reed-Solomon matrix, 
with asymptotically the fastest polarization for $\ell\le q$.

The organization of this paper is as follows.
In Section~\ref{sec:prelim}, notations and definitions used in this paper are introduced.
In Section~\ref{sec:polar}, the basic transform of a source and polarization phenomenon 
by an $\ell\times\ell$ matrix over a finite field
are introduced.
In Section~\ref{sec:equiv}, an equivalence relation of $q$-ary source is defined for
showing equivalence among several polarization problems.
On the concept of equivalence among sources, equivalence of matrices is considered as well.
Using the equivalence of matrices, the main theorem of this paper is stated, which is a necessary and sufficient condition of matrix under which
any source or channel is polarized.
In Section~\ref{sec:bhatt}, the Bhattacharyya parameter and its properties are shown.
They are useful for proving the main theorem in Section~\ref{sec:proof} and
speed of the polarization in Section~\ref{sec:speed}.
In Section~\ref{sec:proof}, a proof of the main theorem is shown.
In Section~\ref{sec:speed}, the speed of the polarization for a general $\ell\times\ell$ matrix is proved similarly to the binary case.
In Section~\ref{sec:RS}, the Reed-Solomon matrices are introduced,
which yield asymptotically the fastest polarization in the sense discussed in Section~\ref{sec:speed}.
In Section~\ref{sec:sim}, the quaternary polar codes using a Reed-Solomon matrix are compared numerically with the original binary polar codes.  
Finally, Section~\ref{sec:sum} summarizes the paper.  

\section{Preliminaries}\label{sec:prelim}
Let $p$ be a prime number and $q := p^m$ where $m$ is a natural number.
Let $\mathbb{F}_q$ be a finite field of size $q$.
Let $\mathbb{F}_q^\times$ be $\mathbb{F}_q\setminus\{0\}$ and $\mathbb{F}_p(\gamma)$ be the simple extension of $\mathbb{F}_p$
generated by the adjunction of $\gamma\in\mathbb{F}_q$.
Similarly, for $A\subseteq\mathbb{F}_q$ and a matrix $G$ over $\mathbb{F}_q$, $\mathbb{F}_p(A)$ and $\mathbb{F}_p(G)$ denote
the field extensions of $\mathbb{F}_p$ generated by the adjunction of all elements of $A$ and $G$, respectively.
Let $\Delta_q:=\{[p_1,\dotsc,p_q]\in\mathbb{R}_{\ge0}^q\mid p_1+\dotsb+p_q=1 \}$ 
denote the set of all $q$-dimensional probability vectors.  
For random variables $X$ on a finite set $\mathcal{X}$ of size $q$ and $Y$ on a discrete set $\mathcal{Y}$,
entropy $H(X)$ of $X$ and conditional entropy $H(X\mid Y)$ of $X$ conditioned on $Y$ are defined as
\begin{align*}
H(X) &:= -\sum_{x\in\mathcal{X}} P_X(x)\log P_X(x)\\
H(X\mid Y) &:= -\sum_{x\in\mathcal{X}, y\in\mathcal{Y}} P_{X,Y}(x,y)\log P_{X\mid Y}(x\mid y).
\end{align*}
In this paper, the base of the logarithm is assumed to be $q$ unless otherwise stated, 
and hence $H(X)$ and $H(X\mid Y)$ are in $[0,1]$.
If a quantity $A((X, Y))$ determined from $P_{X, Y}$ has the form $\mathbb{E}[f([P_{X\mid Y}(x\mid Y)]_{x\in\mathbb{F}_q})]$
for some $f:\Delta_q\to\mathbb{R}$, where $\mathbb{E}$ denotes the expectation, 
we write it as $A(X\mid Y)$ (Here, $P_{X\mid Y}(x\mid Y)$ means the random variable $g(x, Y)$ where $g(x,y):=P_{X\mid Y}(x\mid y)$). 
It should be noted that 
the arguments in this paper are directly applicable to the case 
where $\mathcal{Y}$ is a continuous alphabet such as $\mathbb{R}$, 
by replacing the summation $\sum_{y\in\mathcal{Y}}$ 
with the integral $\int_{-\infty}^{+\infty}\mathrm{d}y$.  
The notation $u_0^{\ell-1}$ denotes the row vector $[u_0, u_1,\dotsc, u_{\ell-1}]$.

\section{Source and channel polarization}\label{sec:polar}
\subsection{Source and channel polarization phenomenon}
In this paper, we consider source polarization on an $\ell\times\ell$ invertible matrix $G$ over $\mathbb{F}_q$.
Let a $q$-ary source $(X, Y)$ be defined as a pair of random variables on $\mathbb{F}_q\times\mathcal{Y}$.
We first introduce a \textit{basic transform} of source, $(X,Y)\to \{(X^{(i)}, Y^{(i)})\}_{i=0,\dotsc,\ell-1}$. 
\begin{definition}[Basic transform]
Let $\{(X_i, Y_i)\}_{i=0,\dotsc,\ell-1}$ be $\ell$ independent drawings of $(X, Y)$.
Let $U_0^{\ell-1}$ be a random vector defined by the equation
$X_0^{\ell-1}=U_0^{\ell-1}G$.
Letting
$(X^{(i)}, Y^{(i)}) := (U_i, (U_0^{i-1}, Y_0^{\ell-1}))$ for $i=0,\dotsc,\ell-1$
defines the basic transform $(X,Y)\to \{(X^{(i)}, Y^{(i)})\}_{i=0,\dotsc,\ell-1}$
where the $\sigma(X_0^{\ell-1}, Y_0^{\ell-1})$-measurable random pair $(X^{(i)}, Y^{(i)})$ 
takes values in $\mathbb{F}_q\times(\mathbb{F}_q^{i}\times\mathcal{Y}^{\ell})$.
\end{definition}
From the chain rule for the entropy, one has 
\begin{align}
&\ell H(X\mid Y) = H(X_0^{\ell-1}\mid Y_0^{\ell-1})= H(U_0^{\ell-1}\mid Y_0^{\ell-1})\nonumber\\
&\quad = \sum_{i=0}^{\ell-1} H(U_i\mid U_0^{i-1}, Y_0^{\ell-1})
 = \sum_{i=0}^{\ell-1} H(X^{(i)}\mid Y^{(i)}).
\label{eq:chain}
\end{align}

By starting with a source $(X, Y)$ and recursively applying the basic transform to depth $n$, we obtain
$\ell^n$ random pairs
$\{(X^{(b_1)\dotsm(b_n)}, Y^{(b_1)\dotsm(b_n)})\}_{(b_1,\dotsc,b_n)\in\{0,\dotsc,\ell-1\}^n}$\footnote{%
Joint distribution of these random pairs is not considered in this paper.}. 
Let $B_1,\dotsc,B_n,\dotsc$ be independent uniform random variables on $\{0,\dotsc,\ell-1\}$.
The random process $\{(\mathsf{X}_n, \mathsf{Y}_n)\}_{n=0,1,\dotsc}$ 
defined via the recursive applications of the basic transform 
as follows puts the foundation of whatever will be discussed in this paper.
\begin{definition}
Let $(\mathsf{X}_n, \mathsf{Y}_n):= (X^{(B_1)\dotsm(B_n)}, Y^{(B_1)\dotsm(B_n)})$
be a $\sigma(X_0^{\ell^n-1}, Y_0^{\ell^n-1}, B_1,\dotsc,B_n)$-measurable random variable
for $n\in\{0,1,\dotsc\}$.
\end{definition}
A random sequence $\{H_n\colon \sigma(B_1,\dotsc,B_n)\text{-measurable} \}_{n=0,1,\dotsc}$ is defined as 
$H_n :=  H(\mathsf{X}_n\mid \mathsf{Y}_n)$ where the conditional entropy does not
take account of randomness of $(B_1,\dotsc,B_n)$.
From the chain rule \eqref{eq:chain} for the entropy, the random sequence $\{H_n\}_{n=0,1,\ldots}$ is shown to be a martingale i.e., $\mathbb{E}[H_n\mid B_1,\dotsc,B_{n-1}] = H_{n-1}$.
Then, noting that the sequence $\{H_n\}_{n=0,1,\ldots}$ is bounded in the interval $[0,1]$, 
from the martingale convergence theorem, there exists a random variable $H_\infty$ such that $H_n$ converges to $H_\infty$ almost surely.
The source polarization is defined in terms of $H_\infty$ 
as in the following definition.
\begin{definition}[Polarization]\label{def:polarization}
A source $(X,Y)$ is said to be polarized by $G$ if and only if
\begin{equation*}
H_\infty = \begin{cases}
0,&\text{\rm with probability } 1-H(X\mid Y)\\
1,&\text{\rm with probability } H(X\mid Y).
\end{cases}
\end{equation*}
\end{definition}
It should be noted that if $H_\infty$ is $\{0,1\}$-valued, 
the probability of $H_\infty=1$ is necessarily equal to $H(X\mid Y)$ 
because of the martingale property $\mathbb{E}[H_n\mid H_0]=H_0=H(X\mid Y)$. 
Note also that Park and Barg~\cite{park2013polar} have adopted a different, 
weaker definition of polarization, in which $H_\infty$ may take more than 
two values. 
In this paper, such cases are regarded as not being polarized.  

When the marginal distribution of $X$ is uniform, the source polarization is called the channel polarization.
As shown in Section~\ref{sec:equiv}, the source polarization problem is also translated into the channel polarization problem.
We therefore use the terms ``source'' and ``channel'' almost interchangeably, 
unless otherwise stated.  
As the first and main contribution of this paper, we show a necessary and sufficient condition of $G$ under which any source or channel is polarized.
Let $G_\gamma:=\begin{bmatrix}1&0\\1&\gamma\end{bmatrix}$ over $\mathbb{F}_q$ where $\gamma\in\mathbb{F}_q^\times$.
Ar{\i}kan proved for the case $q=2$ that the matrix $G_1$ 
polarizes any source/channel~\cite{5075875}, \cite{arikan2010source}.
\sasoglu~et~al.\ generalized the result for prime fields~\cite{sasoglu2009pad}.
They also showed that for the matrix $G_1$ over the ring $\mathbb{Z}/q\mathbb{Z}$ where $q$ is not a prime, there is a counterexample
of non-polarizing $q$-ary channel.
Their counterexample also works for $\mathbb{F}_q$ whose size $q$ is not a prime.
A purpose of this paper is to generalize these results to any matrix over any finite field.

\subsection{Construction of source and channel codes}
\label{sec:construction}
The polar code for source/channel coding is based on the polarization phenomenon.
In this subsection, a rough sketch of construction of the polar code for channel coding is described.
Given an $\ell\times\ell$ invertible matrix $G$ which appears in the previous section,
we first consider an $\ell^n\times \ell^n$ matrix $G^{\otimes n}$
where ${}^{\otimes n}$ denotes the Kronecker power.
For $i\in\{0,1,\dotsc,\ell^n-1\}$, $i_ni_{n-1}\dotsb i_1$ denotes the $\ell$-ary expansion of $i$.
Then, the generator matrix of a polar code is, roughly speaking, obtained from $G^{\otimes n}$ by choosing rows with indices\footnote{%
Row and column indices of matrices start with 0 rather than 1.} in the set 
\begin{equation*}
\{i\in\{0,\dotsc,\ell^n-1\}\mid H(X^{(i_1)\dotsm(i_n)}\mid Y^{(i_1)\dotsm(i_n)}) < \epsilon\}
\end{equation*}
with some threshold $\epsilon >0$.
If a channel $(X, Y)$ is polarized by $G$, the ratio of chosen rows is asymptotically $1-H(X\mid Y)$ for any fixed $\epsilon\in(0,1)$.
For detailed descriptions of encoding and decoding algorithms, see~\cite{5075875} for the channel coding and~\cite{arikan2010source} and \cite{cronie2010lossless} for the source coding.

\section{Equivalence relation on sources and main theorem}\label{sec:equiv}
In order to deal with a source $(X, Y)$ in terms of polarization phenomenon,
it is useful to define an equivalence relation up to which we do not have to distinguish sources.
An equivalence relation $(X, Y) \sim (X', Y')$ which is desirable for our purpose
has to satisfy the following two conditions.
\begin{align}
(X, Y)\sim (X', Y') &\Longrightarrow H(X\mid Y) = H(X'\mid Y')\label{eq:cond1}\\
(X, Y)\sim (X', Y') &\Longrightarrow (X^{(i)}, Y^{(i)}) \sim (X'^{(i)}, Y'^{(i)})\nonumber\\
&\qquad \text{ for } i = 0, 1, \dotsc, \ell-1\label{eq:cond2}
\end{align}
The second condition~\eqref{eq:cond2} should be satisfied 
for any $\ell\times\ell$ invertible matrix $G$.
The significance of these two conditions is that 
sources which are equivalent in the above sense 
yield the same random sequence $\{H_n\}_{n=0,1,\dotsc}$, 
thereby behaving exactly the same as for the polarization phenomenon.  

Given a source $(X, Y)$, the a posteriori distribution 
$[P_{X\mid Y}(x\mid y)]_{x\in\mathbb{F}_q}\in\Delta_q$ plays a fundamental role, 
in particular in determining the conditional entropy $H(X\mid Y)$ 
and other relevant quantities.  
We first introduce two equivalence relations on probability vectors.  
\begin{definition}
For $p_0^{q-1}\in\Delta_q$ and ${p'}_0^{q-1}\in\Delta_q$,
we say $p_0^{q-1}\overset{\mathrm{p}}{\sim}{p'}_0^{q-1}$ if and only if there exists a permutation matrix $\sigma$
such that $p_0^{q-1}={p'}_0^{q-1}\sigma$.
For any $s\in\mathbb{N}$, $[p_{\bm{x}}]_{\bm{x}\in\mathbb{F}_q^s}\in\Delta_{q^s}$ and $[p'_{\bm{x}}]_{\bm{x}\in\mathbb{F}_q^s}\in\Delta_{q^s}$,
we say $[p_{\bm{x}}]_{\bm{x}\in\mathbb{F}_q^s}\overset{\mathrm{q}(s)}{\sim}[p'_{\bm{x}}]_{\bm{x}\in\mathbb{F}_q^s}$ if and only if
there exists $\bm{z}\in\mathbb{F}_q^s$
such that $p_{\bm{x}} = p'_{\bm{x}+\bm{z}}$ for all $\bm{x}\in\mathbb{F}_q^s$.
\end{definition}
It is straightforward to see that 
\begin{equation}\label{eq:equivq}
[p_{\bm{x}}]_{\bm{x}\in\mathbb{F}_q^s}\overset{\mathrm{q}(s)}{\sim}[p'_{\bm{x}}]_{\bm{x}\in\mathbb{F}_q^s}\iff
[p_{\bm{x}H}]_{\bm{x}\in\mathbb{F}_q^s}\overset{\mathrm{q}(s)}{\sim}[p'_{\bm{x}H}]_{\bm{x}\in\mathbb{F}_q^s}
\end{equation}
holds for any $s\times s$ invertible matrix $H$
since $p_{\bm{x}} = p'_{\bm{x}+\bm{z}} \iff p_{\bm{x}H} = p'_{\bm{x}H+\bm{z}H}$ for any $\bm{z}\in\mathbb{F}_q^s$.

The $q$-dimensional random vector 
$[P_{X\mid Y}(x\mid Y)]_{x\in\mathbb{F}_q}\in\Delta_q$ 
induces a probability measure on $\Delta_q$.  
If two random vectors $[P_{X\mid Y}(x\mid Y)]_{x\in\mathbb{F}_q}$ 
and $[P_{X'\mid Y'}(x\mid Y')]_{x\in\mathbb{F}_q}$ defined 
from sources $(X, Y)$ on $\mathbb{F}_q\times\mathcal{Y}$ 
and $(X', Y')$ on $\mathbb{F}_q\times\mathcal{Y}'$, respectively, 
induce the same probability measure on $\Delta_q$, 
we say  $(X, Y)\overset{\mathrm{i}}{\sim}(X',Y')$.
In this case, $A(X\mid Y)=A(X'\mid Y')$ holds for any quantity of the form $A(X\mid Y)
=\mathbb{E}[f([P_{X\mid Y}(x\mid Y)]_{x\in\mathbb{F}_q})]$, and hence the condition~\eqref{eq:cond1} is satisfied.
Furthermore, the equivalence relation $\overset{\mathrm{i}}{\sim}$ obviously satisfies~\eqref{eq:cond2}.
However, a weaker equivalence relation than $\overset{\mathrm{i}}{\sim}$ exists
which satisfies both of the conditions~\eqref{eq:cond1} and~\eqref{eq:cond2}.
First, a weak equivalence relation which only satisfies the condition~\eqref{eq:cond1} is defined as follows.
\begin{definition}
For sources $(X, Y)$ on $\mathbb{F}_q\times\mathcal{Y}$ and $(X', Y')$ on $\mathbb{F}_q\times\mathcal{Y}'$,
we say $(X, Y) \overset{\mathrm{s}}{\sim} (X', Y')$ if and only if
the $q$-dimensional random vector $[P_{X\mid Y}(x\mid Y)]_{x\in\mathbb{F}_q}$ 
induces the same distribution 
on $\Delta_q/\!\overset{\mathrm{p}}{\sim}$ 
as the random vector $[P_{X'\mid Y'}(x\mid Y')]_{x\in\mathbb{F}_q}$.
For a function $f: \Delta_q\to \mathbb{R}$ which is invariant under any permutation of its arguments,
a quantity $\mathbb{E}[f([P_{X\mid Y}(x\mid Y)]_{x\in\mathbb{F}_q})]$
is said to be \statq.
\end{definition}
The equivalence $(X,Y)\overset{\mathrm{s}}{\sim} (X', Y')$ implies 
$A(X\mid Y)=A(X'\mid Y')$ for any quantity $A(X\mid Y)$ \statq, 
including the conditional entropy $H(X\mid Y)$. 
Hence, the equivalence relation $\overset{\mathrm{s}}{\sim}$ satisfies the first condition~\eqref{eq:cond1}.
However, the equivalence relation $\overset{\mathrm{s}}{\sim}$ does not satisfy
the second condition~\eqref{eq:cond2}.
The equivalence relation $\overset{\mathrm{a}(s)}{\sim}$ defined in the following
is weaker than $\overset{\mathrm{i}}{\sim}$ and satisfies
both of the conditions~\eqref{eq:cond1} and~\eqref{eq:cond2}. 
It plays an essential role in the following argument. 
\begin{definition}\label{def:equiva}
Let $s\in\mathbb{N}$.
For pairs of random variables $(X, Y)$ on $\mathbb{F}_q^s\times\mathcal{Y}$ and $(X', Y')$ on $\mathbb{F}_q^s\times\mathcal{Y}'$,
we say $(X, Y) \overset{\mathrm{a}(s)}{\sim} (X', Y')$ if and only if
there exists $r\in\mathbb{F}_q^\times$
such that the $q^s$-dimensional random vector
$[P_{X\mid Y}(r\bm{x}\mid Y)]_{\bm{x}\in\mathbb{F}_q^s}$ induces the same distribution on $\Delta_{q^s}/\!\overset{\mathrm{q}(s)}{\sim}$ 
as $[P_{X'\mid Y'}(\bm{x}\mid Y')]_{\bm{x}\in\mathbb{F}_q^s}$.
\end{definition}
It is not hard to confirm the properties $(X, Y) \overset{\mathrm{i}}{\sim} (X', Y')\Longrightarrow (X, Y) \overset{\mathrm{a}(1)}{\sim} (X', Y')$
and $(X, Y) \overset{\mathrm{a}(1)}{\sim} (X', Y')\Longrightarrow (X, Y) \overset{\mathrm{s}}{\sim} (X', Y')$.
From the latter property, it holds that $(X, Y) \overset{\mathrm{a}(1)}{\sim} (X', Y')\Longrightarrow H(X\mid Y) = H(X'\mid Y')$, 
implying that the equivalence relation $\overset{\mathrm{a}(1)}{\sim}$ 
satisfies the first condition~\eqref{eq:cond1}.  
The equivalence relation $\overset{\mathrm{a}(1)}{\sim}$ also satisfies the second condition~\eqref{eq:cond2}.
\begin{lemma}
\begin{equation*}
(X, Y)\overset{\mathrm{a}(1)}{\sim} (X', Y') \Longrightarrow (X^{(i)}, Y^{(i)}) \overset{\mathrm{a}(1)}{\sim} (X'^{(i)}, Y'^{(i)})
\end{equation*}
for $i = 0, 1, \dotsc, \ell-1$ and
for an arbitrary $\ell\times\ell$ invertible matrix $G$. 
\end{lemma}
\begin{IEEEproof}
For a source $(X, Y)$,
let $X_0^{\ell-1}$, $Y_0^{\ell-1}$ and $U_0^{\ell-1}$ be what appear in the definition of the basic transform of it.
The random variables
${X'}_0^{\ell-1}$, ${Y'}_0^{\ell-1}$ and ${U'}_0^{\ell-1}$ are defined in the same way for $(X', Y')$.
The equivalence relation $(X, Y)\overset{\mathrm{a}(1)}{\sim}(X',Y')$ 
between sources $(X, Y)$ and $(X', Y')$ 
immediately leads to the equivalence 
$(X_0^{\ell-1}, Y_0^{\ell-1})\overset{\mathrm{a}(\ell)}{\sim}({X'}_0^{\ell-1},{Y'}_0^{\ell-1})$ between their $\ell$th-order extensions.
From~\eqref{eq:equivq} and the identity $(r\bm{x})G^{-1}=r(\bm{x}G^{-1})$ for any $r\in\mathbb{F}_q^\times$ and $\bm{x}\in\mathbb{F}_q^\ell$,
it holds that $(X_0^{\ell-1}G^{-1}, Y_0^{\ell-1})\overset{\mathrm{a}(\ell)}{\sim}({X'}_0^{\ell-1}G^{-1}, {Y'}_0^{\ell-1})$, or equivalently,  
$(U_0^{\ell-1}, Y_0^{\ell-1})\overset{\mathrm{a}(\ell)}{\sim}({U'}_0^{\ell-1}, {Y'}_0^{\ell-1})$.  
One therefore obtains $(U_i, (U_0^{i-1}, Y_0^{\ell-1}))\overset{\mathrm{a}(1)}{\sim}({U'}_i, ({U'}_0^{i-1}, {Y'}_0^{\ell-1}))$.
\end{IEEEproof}
The equivalence relation $\overset{\mathrm{a}(1)}{\sim}$ gives rise to the following several useful lemmas.

\begin{lemma}[Source-channel equivalence~\cite{hussami2009performance}]\label{lem:s-c}
Let $(N, Z)$ be a random pair on $\mathbb{F}_q\times\mathcal{Y}$ and
$X$ be a uniform random variable on $\mathbb{F}_q$ which is independent of $(N,Z)$. 
Then, it holds that $(N, Z)\overset{\mathrm{a}(1)}{\sim} (X, (X+N,Z))$.
\end{lemma}
\begin{IEEEproof}
One has 
$(X, (X+N,Z))\overset{\mathrm{a}(1)}{\sim}(-X + (X+N), (X+N,Z))=
(N, (X+N,Z))\overset{\mathrm{i}}{\sim}(N, Z)$, where the last 
equivalence relation is due to the assumptions on $X$.  
\end{IEEEproof}
The channel $(X, (X+N, Z))$ in Lemma~\ref{lem:s-c} is a symmetric channel in the following sense.
\begin{definition}[Symmetric channel]
A channel $(X, Y)$ on $\mathbb{F}_q\times \mathcal{Y}$ is said to be symmetric if and only if
there exists a permutation $\sigma_x$ on $\mathcal{Y}$ for each $x\in\mathbb{F}_q$ such that
$P_{Y\mid X}(y\mid x)=P_{Y\mid X}(\sigma_{x'-x}(y)\mid x')$ for any $y\in\mathcal{Y}$ and $x, x'\in\mathbb{F}_q$.
\end{definition}
The symmetricity is preserved under the basic transform.
\begin{lemma}\label{lem:psym}
For a symmetric channel $(X, Y)$, 
$(X^{(i)}, Y^{(i)})$ is symmetric for any $i\in\{0,\dotsc,\ell-1\}$.
\end{lemma}
\begin{IEEEproof}
The statement holds since
$P_{U_0^{\ell-1}, Y_0^{\ell-1}}((u_0^{i-1},u_i,u_{i+1}^{\ell-1}), y_0^{\ell-1})=
P_{U_0^{\ell-1}, Y_0^{\ell-1}}((u_0^{i-1},u_i',u_{i+1}^{\ell-1}), w_0^{\ell-1})$
where $w_j=\sigma_{G_{i,j}(u'_i-u_i)}(y_j)$.
\end{IEEEproof}

The following technical lemma implies that one can ignore effects 
of addition of a known constant to input of symmetric channels.  
It will be used in the proof of the main theorem. 
\begin{lemma}\label{lem:sym}
For any channel $(X, Y)$ and any symmetric channel $(X', Y')$, 
let $(Z, (Y, Y'))$ and $(Z', (Y, Y'))$ be the channels 
defined by letting $Z=X=X'$ and $Z'=X=X'+a$ for any fixed $a\in\mathbb{F}_q$, 
respectively.  
For these channels, 
it holds that $(Z, (Y, Y'))\overset{\mathrm{i}}{\sim} (Z', (Y, Y'))$.
\end{lemma}
\begin{IEEEproof}
The equality
$P_{Z,(Y,Y')}(z,(y,y'))=P_{Z',(Y,Y')}(z,(y,\sigma_a^{-1}(y')))$
implies $(Z,(Y,Y')) \overset{\mathrm{i}}{\sim} (Z', (Y, Y'))$.
\end{IEEEproof}

We next introduce an equivalence relation on matrices on the basis of 
the equivalence relation $\overset{\mathrm{a}(1)}{\sim}$ on sources/channels.  
We say that $\ell\times\ell$ invertible matrices $\hat{G}$ and $\bar{G}$ are equivalent when
$(\hat{X}^{(i)}, \hat{Y}^{(i)})\overset{\mathrm{a}(1)}{\sim}(\bar{X}^{(i)}, \bar{Y}^{(i)})$ for $i=0,\dotsc,\ell-1$
where $\{(\hat{X}^{(i)}, \hat{Y}^{(i)})\}_{i=0,\dotsc,\ell-1}$ and $\{(\bar{X}^{(i)}, \bar{Y}^{(i)})\}_{i=0,\dotsc,\ell-1}$ 
are two sets of $\ell$ random pairs generated from an arbitrary common source $(X, Y)$
via the basic transform using matrices $\hat{G}$ and $\bar{G}$, respectively.

\begin{lemma}
\label{lem:equivM}
Let $G$ and $V$ be an $\ell\times\ell$ invertible matrix 
and an $\ell\times\ell$ invertible upper triangular matrix, respectively. 
Then, $G$ and $VG$ are equivalent.
\end{lemma}
\begin{IEEEproof}
Since $X_0^{\ell-1} = U_0^{\ell-1}VG \iff X_0^{\ell-1} G^{-1}= U_0^{\ell-1}V =: {U'}_0^{\ell-1}$,
the equivalence $(U_i, (U_0^{i-1}, Y_0^{\ell-1}))\overset{\mathrm{a}(1)}{\sim} (U'_i, ({U'}_0^{i-1}, Y_0^{\ell-1}))$
implies the lemma.
\end{IEEEproof}
Obviously, a permutation of columns of $G$ does not change $(X^{(i)}, Y^{(i)})$ up to the equivalence $\overset{\mathrm{i}}{\sim}$
for $i = 0,\dotsc,\ell-1$, 
so that $G$ and its column permutation are equivalent.  
Hence, without loss of generality, one can assume that $G$ is a lower triangular matrix.
\begin{definition}[Standard form]
Lower triangular matrices with unit diagonal elements equivalent to $G$ are called standard forms of $G$.
\end{definition}
A standard form of $G$ is not generally unique.
For example, the standard forms of $G_\gamma$ are 
$\begin{bmatrix}1&0\\\gamma^{-1}&1\end{bmatrix}$ and
$\begin{bmatrix}1&0\\\gamma&1\end{bmatrix}$.
If there exists the identity matrix as a standard form of $G$,
it is the unique standard form of $G$.
In this case, one obviously has the identity $(X^{(i)}, Y^{(i)})\overset{\mathrm{a}(1)}{\sim}(X, Y)$ for all $i\in\{0,\dotsc,\ell-1\}$, 
implying that $G$ does not polarize any source.  
For other cases, the following main theorem shows necessary and sufficient conditions of $G$ under which any source is polarized.

\begin{theorem}\label{thm:main}
The followings are equivalent for an $\ell\times\ell$ invertible matrix $G$ over $\mathbb{F}_q$ with a non-identity standard form.
\begin{itemize}
\item Any $q$-ary source is polarized by $G$.
\item It holds $\mathbb{F}_p(\bar{G})=\mathbb{F}_q$ for any standard form $\bar{G}$ of $G$.
\item It holds $\mathbb{F}_p(\bar{G})=\mathbb{F}_q$ for one of the standard forms $\bar{G}$ of $G$.
\end{itemize}
\end{theorem}

\begin{corollary}
Any $q$-ary source is polarized by the $2\times 2$ matrix $G_\gamma$ over $\mathbb{F}_q$ with $\gamma\in\mathbb{F}_q^\times$ if and only if
$\mathbb{F}_p(\gamma)=\mathbb{F}_q$.
\end{corollary}

Note that the identity matrix is the standard form of an invertible matrix $G$ 
if and only if there exists an upper triangular matrix as a column permutation of $G$.
Thus, Theorem~\ref{thm:main} includes the known results 
that an invertible matrix $G$ is polarizing 
if and only if any column permutation of $G$ is not upper triangular 
for $q=2$~\cite[Lemma 1]{korada2010polar} 
and for $q$ prime~\cite{sasoglu2009pad}.

\section{Bhattacharyya parameter}\label{sec:bhatt}
Bhattacharyya parameter is useful both for proving the polarization phenomenon, and for evaluating asymptotic speed of polarization.
In this section, it is shown that polarization of Bhattacharyya parameter and polarization of the conditional entropy are equivalent.
Let $(\Omega := \{1,\dotsc,q\}, 2^{\Omega}, P)$ be a probability space.
The probability measure $P$ can be represented by the vector $[\sqrt{P(1)},\dotsc,\sqrt{P(q)}]\in \mathcal{S}_q$
where $\mathcal{S}_q:=\{[p_1,\dotsc,p_q]\in\mathbb{R}_{\ge0}^q\mid p_1^2+\dotsb+p_q^2=1\}$.
The $L_p$ norm of $\bm{x}\in\mathbb{C}^q$ is defined as $L_p(\bm{x}) := (|x_1|^p+\dotsb+|x_q|^p)^{1/p}$ for any $p\ge 1$.
The $L_1$ norm of $\bm{p}\in \mathcal{S}_q$ attains the minimum 1 at the deterministic distributions i.e., the distributions of the form $[0,\dotsc,0,1,0,\dotsc,0]$,
and the maximum $\sqrt{q}$ at the uniform distribution, represented by 
$\bm{u}:=[1/\sqrt{q},\dotsc,1/\sqrt{q}]\in\mathcal{S}_q$.
On the other hand, the deterministic and uniform distributions also minimize and maximize the entropy
$H(\bm{p}) := -\sum_i p_i^2\log p_i^2$ of $\bm{p}\in \mathcal{S}_q$, respectively. 

The following lemma states that closeness of a probability distribution 
to determinism or uniformity measured in terms of its entropy value 
is equivalent to that measured in terms of its $L_1$-norm value. 

\begin{lemma}\label{lem:HZ}
For any $\epsilon>0$, there exists $\delta>0$
such that
\begin{align}
\label{eq:HL1}\{\bm{p}\in \mathcal{S}_q\mid H(\bm{p})<\delta\}&\subseteq\{\bm{p}\in \mathcal{S}_q\mid L_1(\bm{p})-1<\epsilon\}\\
\label{eq:HL2}\{\bm{p}\in \mathcal{S}_q\mid L_1(\bm{p})-1<\delta\}&\subseteq\{\bm{p}\in \mathcal{S}_q\mid H(\bm{p})<\epsilon\}\\
\label{eq:HL3}\{\bm{p}\in \mathcal{S}_q\mid 1-H(\bm{p})<\delta\}&\subseteq\{\bm{p}\in \mathcal{S}_q\mid \sqrt{q}-L_1(\bm{p})<\epsilon\}\\
\label{eq:HL4}\{\bm{p}\in \mathcal{S}_q\mid \sqrt{q}-L_1(\bm{p})<\delta\}&\subseteq\{\bm{p}\in \mathcal{S}_q\mid 1-H(\bm{p})<\epsilon\}.
\end{align}

\end{lemma}
\begin{IEEEproof}
Since
\begin{align*}
L_2(\bm{u}-\bm{p})^2 &= \sum_{i=1}^q \left(\frac1{\sqrt{q}} - p_i\right)^2
= 2-\frac2{\sqrt{q}}\sum_{i=1}^q p_i\\
&= \frac2{\sqrt{q}}\left(\sqrt{q}-L_1(\bm{p})\right)
\end{align*}
\eqref{eq:HL4} is a consequence of continuity of $H(\bm{p})$.
The relationship \eqref{eq:HL3} follows from
\begin{align*}
&1-H(\bm{p})=1+\sum_{i=1}^qp_i^2\log p_i^2 =
-2\sum_{i=1}^qp_i^2\log \frac1{\sqrt{q} p_i}\\
&\,\ge \frac{2}{\log_{\mathrm{e}} q}\sum_{i=1}^qp_i^2\left(1-\frac1{\sqrt{q} p_i}\right)
= \frac{2}{\sqrt{q}\log_{\mathrm{e}} q}\left(\sqrt{q}-L_1(\bm{p})\right).
\end{align*}

Since $H(\bm{p})=2\sum_{i}p_i^2\log (1/p_i) \le 2\log\sum_{i}p_i=2\log L_1(\bm{p})$,
the relationship \eqref{eq:HL2} holds.
Since $H(\bm{p})\log_{\mathrm{e}} q =-\sum_{i}p_i^2\log_{\mathrm{e}} p_i^2 \ge -\log_{\mathrm{e}} \max_i p_i^2 \ge 1-\max_i p_i^2
\ge (L_1(\bm{p})-1)^2/(q-1)$
(see \eqref{eq:l1max} for the last inequality),
the relationship \eqref{eq:HL1} holds.
\end{IEEEproof}
Hence, the entropy is close to 0 and 1 if and only if the $L_1$ norm is close to 1 and $\sqrt{q}$, respectively.

The above argument is applied to random pairs $(X,Y)$ to establish 
the relationship between the conditional entropy and Bhattacharyya parameter. 
The expectation of the squared $L_1$ norm of the a posteriori probability 
vector $[\sqrt{P_{X\mid Y}(x\mid Y)}]_{x\in\mathbb{F}_q}\in\mathcal{S}_q$ satisfies
\begin{align}
&1\le \sum_{y\in\mathcal{Y}} P_Y(y)\left[\sum_{x\in\mathbb{F}_q}\sqrt{P_{X\mid Y}(x\mid y)}\right]^2\le q\nonumber\\
\Longleftrightarrow\;&  \nonumber\\
0\le&\frac1{q-1}\sum_{\substack{x\in\mathbb{F}_q, x'\in\mathbb{F}_q\\ x\ne x'}}\sum_{y\in\mathcal{Y}}
P_Y(y)\sqrt{P_{X\mid Y}(x\mid y)P_{X\mid Y}(x'\mid y)}\nonumber\\
&\le 1
\label{eq:ineqL1}
\end{align}
for any random pair $(X,Y)$. 
From Lemma~\ref{lem:HZ} and~\eqref{eq:ineqL1},
the conditional entropy $H(X\mid Y)$ is close to 0 and 1 if and only if
the \textit{Bhattacharyya parameter} $Z(X\mid Y)\in[0,1]$ for $(X,Y)$, 
defined as follows, is close to 0 and 1, respectively.
\begin{definition}[Bhattacharyya parameter]
\begin{align*}
&Z(X\mid Y) \nonumber\\
&:=\frac1{q-1}\sum_{\substack{x\in\mathbb{F}_q, x'\in\mathbb{F}_q\\ x\ne x'}}
\sum_{y\in\mathcal{Y}}P_Y(y)\sqrt{P_{X\mid Y}(x\mid y)P_{X\mid Y}(x'\mid y)}.
\end{align*}
\end{definition}
Obviously, $Z(X\mid Y)$ is \statq\ of $(X, Y)$.
For $d\in\mathbb{F}_q^\times$, we define $Z_d(X\mid Y)\in[0,1]$ as
\begin{equation*}
Z_d(X\mid Y) :=\sum_{x\in\mathbb{F}_q}\sum_{y\in\mathcal{Y}}P_Y(y)\sqrt{P_{X\mid Y}(x\mid y)P_{X\mid Y}(x+d\mid y)}.
\end{equation*}
The Bhattacharyya parameter $Z(X\mid Y)$ can be expressed as the average of $Z_d(X\mid Y)$
\begin{equation*}
Z(X\mid Y)=\frac1{q-1}\sum_{d\in\mathbb{F}_q^\times} Z_d(X\mid Y).
\end{equation*}
Hence, $Z(X\mid Y)$ is close to 0 and 1 if and only if $Z_d(X\mid Y)$ is simultaneously close to 0 and 1 for all $d\in\mathbb{F}_q^\times$, respectively.

\section{Proof of the main theorem}\label{sec:proof}
\subsection{Sketch}
In this section, the proof of Theorem~\ref{thm:main} is shown.
In Section~\ref{subsec:necessity},
it is proved that if there exists a standard form $\bar{G}$ of $G$ such that $\mathbb{F}_p(\bar{G})\ne\mathbb{F}_q$,
there exists a source which is not polarized by $G$.
It means that if any source is polarized by $G$, any standard form $\bar{G}$ of $G$ satisfies $\mathbb{F}_p(\bar{G})=\mathbb{F}_q$.
In Section~\ref{subsec:sufficiency},
it is proved that if there exists a standard form $\bar{G}$ of $G$ such that $\mathbb{F}_p(\bar{G})=\mathbb{F}_q$,
any source is polarized by $G$.
This completes the proof of Theorem~\ref{thm:main}.

\subsection{Necessity}\label{subsec:necessity}
Let $\bar{G}$ be an arbitrary standard form of $G$.
Assume $\mathbb{F}_p(\bar{G}) \ne \mathbb{F}_q$.
Let $M := [\mathbb{F}_q : \mathbb{F}_p(\bar{G})]$ be a degree of a field extension $\mathbb{F}_q/\mathbb{F}_p(\bar{G})$.
Since $\mathbb{F}_q/\mathbb{F}_p(\bar{G})$ is an $M$-dimensional linear space over $\mathbb{F}_p(\bar{G})$,
there is an isomorphism $\psi\colon \mathbb{F}_q/\mathbb{F}_p(\bar{G})\to \mathbb{F}_p(\bar{G})^M$.
Let $[V_0,\dotsc,V_{M-1}] \in \mathbb{F}_p(\bar{G})^M$ be the random vector $\psi(X)$ for
$X\in\mathbb{F}_q$.
If one takes a source $(X,Y)$ for which $V_0,\dotsc,V_{M-1}$ are independent conditioned on $Y$,
recursive application of the basic transform to the source $(X, Y)$ affects $V_i$ separately for $i\in\{0,\dotsc,M-1\}$,
i.e.,
one can regard the polarization process of the source $(X, Y)$ 
as a collection of $M$ distinct polarization processes 
$\{(\mathsf{V}_{i,n},\mathsf{Y}_n) :=(V_i^{(B_1)\cdots(B_n)},Y^{(B_1)\cdots(B_n)})\}_{n=0,1,\dotsc}$, 
$i=0,\dotsc,M-1$. 
In this case, if $H(V_i\mid Y)$ is not constant among all $i\in\{0,\dotsc,M-1\}$,
the source $([V_0,\dotsc, V_{M-1}],Y)$ cannot be polarized in principle, 
in the sense defined in Definition~\ref{def:polarization}. 
Note that
the situation is essentially equivalent to the polar coding for the $M$-user multiple access channel~\cite{abbe2012polar}.

\subsection{Sufficiency}\label{subsec:sufficiency}
In the proof of sufficiency, $(X, Y)$ is assumed to be a symmetric channel.
From Lemma~\ref{lem:s-c} we do not lose generality by this assumption.
For any $j\in\{0,\dotsc,\ell-1\}$, it holds via the chain rule for the entropy that
\begin{align}
\sum_{i=j}^{\ell-1} H(X^{(i)}\mid Y^{(i)})
&=H(U_{j}^{\ell-1}\mid U_0^{j-1}, Y_0^{\ell-1})\nonumber\\
&=\sum_{i=j}^{\ell-1} H(U_{i}\mid U_0^{j-1}, U_{i+1}^{\ell-1}, Y_0^{\ell-1})\label{eq:Hsum}
\end{align}
for any $(X, Y)$.
Let $\bar{G}$ be an arbitrary standard form of $G$, 
and assume that $U_0^{\ell-1}$ and $(X^{(i)}, Y^{(i)})$ for $i\in\{0,\dotsc,\ell-1\}$ 
are defined with $\bar{G}$.  
All the terms in the rightmost side of~\eqref{eq:Hsum}
are at most $H(X\mid Y)$ for any standard form $\bar{G}$.  
It also holds that $|H(\mathsf{X}_n^{(i)}\mid \mathsf{Y}_n^{(i)})-H(\mathsf{X}_n\mid \mathsf{Y}_n)|\to 0$ with probability 1
as $n\to\infty$ for all $i\in\{0,\dotsc,\ell-1\}$
since $\{H(\mathsf{X}_n\mid \mathsf{Y}_n)\}_{n=0,1,\ldots}$ 
converges almost surely.
Combining these two facts, one observes that 
each of the terms in the sum on the rightmost side of~\eqref{eq:Hsum}
evaluated with $(X, Y) = (\mathsf{X}_n, \mathsf{Y}_n)$ 
must be close to $H(\mathsf{X}_n\mid \mathsf{Y}_n)$ with probability 1
as $n\to\infty$.
In particular, 
\begin{equation*}
H(\mathsf{X}_n\mid\mathsf{Y}_n) 
- \left.H(U_{j}\mid U_0^{j-1}, U_{j+1}^{\ell-1}, Y_0^{\ell-1})\right|_{(X, Y) = (\mathsf{X}_n, \mathsf{Y}_n)}\to 0
\end{equation*}
holds with probability 1.
Hence, it also holds
\begin{equation}
H(\mathsf{X}_n\mid\mathsf{Y}_n) 
- \left.H(U_{j}\mid U_0^{j-1}, U_{j+1}^{\ell-1}, Y_k, Y_j)\right|_{(X, Y) = (\mathsf{X}_n, \mathsf{Y}_n)}\to 0
\label{eq:up}
\end{equation}
for any $0\le k<j\le \ell-1$.
From Lemmas~\ref{lem:psym} and \ref{lem:sym}, the effects of $U_0^{j-1}$ and $U_{j+1}^{\ell-1}$ can be ignored, i.e., it holds $(U_j, (U_0^{j-1}, U_{j+1}^{\ell-1}, Y_k, Y_j)) \overset{\mathrm{i}}{\sim} (U_j, (Y_k, Y_j))$
where the channel on the right-hand side is defined from 
$(U_j, (U_0^{j-1}, U_{j+1}^{\ell-1}, Y_k, Y_j))$ by fixing 
$U_0^{j-1}$ and $U_{j+1}^{\ell-1}$ to the all-zero vectors.
Assume that the $(j, k)$-element of $\bar{G}$ is $\gamma\ne 0$.
Let $\{(\bar{\mathsf{X}}_n^{(0)}, \bar{\mathsf{Y}}_n^{(0)}), (\bar{\mathsf{X}}_n^{(1)}, \bar{\mathsf{Y}}_n^{(1)})\}$ be the random pairs 
obtained from $(\mathsf{X}_n, \mathsf{Y}_n)$ via the basic transform 
with the $2\times2$ matrix $\begin{bmatrix}1&0\\\gamma&1\end{bmatrix}$, which is a standard form of $G_\gamma$.  
Then, from~\eqref{eq:up},
it holds that $H(\mathsf{X}_n\mid \mathsf{Y}_n)-H(\bar{\mathsf{X}}_n^{(1)}\mid \bar{\mathsf{Y}}_n^{(1)})\to 0$ with probability 1.
The relationships of random variables are described in Fig.~\ref{fig:process}.
In the rest of the proof, we do not use the relationship between $(\mathsf{X}_n, \mathsf{Y}_n)$ and $(\mathsf{X}_{n-1}, \mathsf{Y}_{n-1})$,
and only use the fact that
$H(\mathsf{X}_n\mid \mathsf{Y}_n)-H(\bar{\mathsf{X}}_n^{(1)}\mid \bar{\mathsf{Y}}_n^{(1)})\to 0$ with probability 1 for $G_\gamma$ where $\gamma$ is an arbitrary off-diagonal non-zero element of $\bar{G}$.
The following proposition implies the sufficiency of the main theorem.
\begin{figure}
\begin{center}
\begin{tikzpicture}
\node at (0,0) {$(\mathsf{X}_0,\mathsf{Y}_0) \overset{G_\gamma}{\to}
\{(\bar{\mathsf{X}}_0^{(0)},\bar{\mathsf{Y}}_0^{(0)}),(\bar{\mathsf{X}}_0^{(1)},\bar{\mathsf{Y}}_0^{(1)})\}$};
\node at (0,-.5) {{\footnotesize $G$}$\downarrow\hspace{13.0em}$};
\node at (0,-1) {$(\mathsf{X}_1,\mathsf{Y}_1) \overset{G_\gamma}{\to}
\{(\bar{\mathsf{X}}_1^{(0)},\bar{\mathsf{Y}}_1^{(0)}),(\bar{\mathsf{X}}_1^{(1)},\bar{\mathsf{Y}}_1^{(1)})\}$};
\node at (0,-1.5) {{\footnotesize $G$}$\downarrow\hspace{13.0em}$};
\node at (0,-2) {$\vdots\hspace{12.5em}$};
\node at (0,-2.5) {$(\mathsf{X}_n,\mathsf{Y}_n) \overset{G_\gamma}{\to}
\{(\bar{\mathsf{X}}_n^{(0)},\bar{\mathsf{Y}}_n^{(0)}),(\bar{\mathsf{X}}_n^{(1)},\bar{\mathsf{Y}}_n^{(1)})\}$};
\node at (0,-3) {{\footnotesize $G$}$\downarrow\hspace{13.0em}$};
\node at (0,-3.5) {$\vdots\hspace{12.5em}$};
\end{tikzpicture}
\caption{
The relationships of $(\mathsf{X}_n,\mathsf{Y}_n)$ and $(\mathsf{X}_n^{(1)},\mathsf{Y}_n^{(1)})$.
In the vertical arrows, the basic transform defined in Section~\ref{sec:polar} based on the matrix $G$ is applied.
In the horizontal arrows, the basic transform based on the matrix $G_\gamma$ is applied.
}
\label{fig:process}
\end{center}
\end{figure}
\begin{proposition}\label{prop:polar}
Let $A$ be a non-empty subset of $\mathbb{F}_q^\times$.
Let $\{(X_{(n)}, Y_{(n)})\}_{n=0,1,\dotsc}$ be a sequence of random pairs.
Assume $H(X_{(n)}\mid Y_{(n)})-H(X_{(n)}^{(1)}\mid Y_{(n)}^{(1)}) \to 0$ for all $G_\gamma$ where $\gamma\in A$.
Then,
for any $\epsilon>0$, there exists $n_0\in\mathbb{N}$ such that
\begin{align*}
Z_{t d}(X_{(n)} \mid Y_{(n)}) &< \epsilon,\hspace{2em} \text{ for all }\hspace{1em} t\in\mathbb{F}_p(A)^\times\\
\hspace{0.5em}
\text{ or }
\hspace{1.5em}
Z_{t d}(X_{(n)} \mid Y_{(n)}) &> 1 - \epsilon,\hspace{2em} \text{ for all }\hspace{1em} t\in\mathbb{F}_p(A)^\times
\end{align*}
for any $n\ge n_0$ and any $d\in\mathbb{F}_q^\times$.
\end{proposition}
When $\mathbb{F}_p(\bar{G}) = \mathbb{F}_q$, 
Proposition~\ref{prop:polar} states that the random sequence 
$H_n=H(\mathsf{X}_n\mid\mathsf{Y}_n)$ is close to 0 or 1 for sufficiently large $n$ with probability 1.
Hence, $H_\infty$ must be $\{0,1\}$-valued, i.e., any source $(X,Y)$ is polarized by $G$.
\noindent

What remains is to prove Proposition~\ref{prop:polar}. 
It is equivalent to the following proposition, which will be proved 
in the rest of this section.  
\begin{proposition}\label{prop:polar2}
Let $A$ be a non-empty subset of $\mathbb{F}_q^\times$.
Let $\{(X_{(n)}, Y_{(n)})\}_{n=0,1,\dotsc}$ be a sequence of random pairs.
Assume $H(X_{(n)}\mid Y_{(n)})-H(X_{(n)}^{(1)}\mid Y_{(n)}^{(1)}) \to 0$ for all $G_\gamma$ where $\gamma\in A$.
Then,
for any $\epsilon>0$, there exists $n_0\in\mathbb{N}$ such that
\begin{itemize}
\item[(p0)] $Z_d(X_{(n)}\mid Y_{(n)})<\epsilon$ or $Z_d(X_{(n)}\mid Y_{(n)})>1-\epsilon$,
\item[(p1)] $\bigl(Z_d(X_{(n)}\mid Y_{(n)})>1-\epsilon \Rightarrow Z_{\gamma d}(X_{(n)}\mid Y_{(n)})>1-\epsilon\bigr)$ for any
$\gamma\in A$,
\item[(p2)] $\bigl(\bigl(Z_d(X_{(n)}\mid Y_{(n)})>1-\epsilon$ and $Z_{d'}(X_{(n)}\mid Y_{(n)})>1-\epsilon\bigr) \Rightarrow Z_{d+d'}(X_{(n)}\mid Y_{(n)})>1-\epsilon\bigr)$
for any $d'\in\mathbb{F}_q^\times$,
\end{itemize}
for any $n\ge n_0$ and any $d\in\mathbb{F}_q^\times$.
\end{proposition}
Note that from (p0), $\gamma^{q-1}=1$ and $\overbrace{d'+d'+\dotsb d'}^{p \text{ times}}=0$, the
conditions (p1) and (p2) imply
\begin{itemize}
\item[(p'1)] $\bigl(Z_d(X_{(n)}\mid Y_{(n)})<\epsilon \Rightarrow Z_{\gamma d}(X_{(n)}\mid Y_{(n)})<\epsilon\bigr)$ for any
$\gamma\in A$,
\item[(p'2)] $\bigl(\bigl(Z_d(X_{(n)}\mid Y_{(n)})<\epsilon$ and $Z_{d'}(X_{(n)}\mid Y_{(n)})<\epsilon\bigr) \Rightarrow Z_{d+d'}(X_{(n)}\mid Y_{(n)})<\epsilon\bigr)$
for any $d'\in\mathbb{F}_q^\times$,
\end{itemize}
respectively,
for any $n\ge n_0$ and any $d\in\mathbb{F}_q^\times$.
It is easy to confirm that Proposition~\ref{prop:polar} implies Proposition~\ref{prop:polar2}.
The other direction also holds since $\mathbb{F}_p(A)=\{\gamma_1^{i_1}+\gamma_2^{i_2}+\dotsb +\gamma_m^{i_m}\mid m\in\mathbb{N}, i_j=0,1,\dotsc,q-2, \gamma_j\in A, \forall j=1,2,\dotsc,m\}$.
\begin{remark}
Note that among the three conditions (p0), (p1) and (p2), only (p1) uses the set $A$.
Indeed, (p0) and (p2) hold for any matrix as shown in~\cite{sasoglu2009pad}.
When $q$ is a prime, the
 conditions (p0) and (p2) are sufficient to prove Proposition~\ref{prop:polar} since $\mathbb{F}_p^{\times}=\{1, 1+1, 1+1+1, \dotsc, \overbrace{1+\dotsb+1}^{p-1 \text{ 1s}}\}$~\cite{sasoglu2009pad}.
When $A$ includes a primitive element $\gamma$ of $\mathbb{F}_p(A)$, i.e., $\mathbb{F}_p(A)^\times=\{1, \gamma, \gamma^2, \dotsc, \gamma^{q-2}\}$, the 
 conditions (p0) and (p1) are also sufficient to prove Proposition~\ref{prop:polar}~\cite{mori2010cpq}.
However, generally, we need all of (p0), (p1) and (p2) for proving Proposition~\ref{prop:polar}.
\end{remark}
The following lemma implies (p0) and (p1) to hold 
under the assumptions of Proposition~\ref{prop:polar2}.
\begin{lemma}\label{lem:Zd}
Let $\{(X_{(n)}, Y_{(n)})\}_{n=0,1,\dotsc}$ be a sequence of random pairs.
Assume $H(X_{(n)}\mid Y_{(n)})-H(X_{(n)}^{(1)}\mid Y_{(n)}^{(1)}) \to 0$ for $G_\gamma$ where $\gamma\in\mathbb{F}_q^\times$.
Then, for any $\epsilon > 0$ there exists $n_0\in\mathbb{N}$ such that
\begin{align*}
Z_{\gamma^i d}(X_{(n)} \mid Y_{(n)}) &< \epsilon,\hspace{1em} \text{ for all }\hspace{0.5em} i=0,\dots,q-2\\
\hspace{0.5em}
\text{ or }
\hspace{0.5em}
Z_{\gamma^i d}(X_{(n)} \mid Y_{(n)}) &> 1 - \epsilon,\hspace{1em} \text{ for all }\hspace{0.5em} i=0,\dots,q-2
\end{align*}
for any $n\ge n_0$ and any $d\in\mathbb{F}_q^\times$.
\end{lemma}
\noindent
The proof of Lemma~\ref{lem:Zd} is in Appendix~\ref{sec:proofZd}.
The following lemma and (p0) imply (p2) to hold 
under the assumptions of Proposition~\ref{prop:polar2}, 
completing the proof of sufficiency of the main theorem.
\begin{lemma}[\cite{sasoglu2009pad}]\label{lem:Ztri}
For any $d_1$ and $d_2$ in $\mathbb{F}_q^\times$ satisfying $d_2\ne -d_1$, 
\begin{align*}
&\sqrt{1-Z_{d_1+d_2}(X\mid Y)}\\
&\qquad\le\sqrt{1-Z_{d_1}(X\mid Y)} + \sqrt{1-Z_{d_2}(X\mid Y)}.
\end{align*}
\end{lemma}
\begin{IEEEproof}
Since
\begin{align*}
&1-Z_{d}(X\mid Y)\\
&\quad = \frac12\sum_{x\in\mathbb{F}_q}\sum_{y\in\mathcal{Y}}\Big(\sqrt{P_{X, Y}(x, y)} - \sqrt{P_{X, Y}(x+d, y)}\Big)^2
\end{align*}
the statement is obtained from the triangle inequality of the Euclidean distance.
\end{IEEEproof}

\section{Error probability, total variation distance to the uniform distribution and speed of polarization}\label{sec:speed}
\subsection{Preliminaries}
In this section, we consider speed of polarization by an $\ell\times\ell$ invertible matrix $G$ over $\mathbb{F}_q$.
Let
\begin{align*}
P_\mathrm{e}(X\mid Y) &:= 1- \sum_{y\in\mathcal{Y}}P_Y(y)\max_{x\in\mathbb{F}_q} P_{X\mid Y}(x\mid y).
\end{align*}
This is the average error probability of the maximum a posteriori estimator
$\hat{x}(y):=\arg\max_{x\in\mathbb{F}_q}P_{X\mid Y}(x\mid y)$ of $X$ given $Y$. 
The random quantity $P_\mathrm{e}(\mathsf{X}_n\mid\mathsf{Y}_n)$ plays 
a key role in studying speed of polarization.  
It provides a bound of the block error probability of polar codes 
with successive cancellation decoding 
applied to channel coding~\cite{5075875}.  
More precisely, if one has 
\begin{equation*}
\Pr(P_\mathrm{e}(\mathsf{X}_n\mid\mathsf{Y}_n)<\epsilon)\ge R
\end{equation*}
then it implies existence of a polar code for channel coding 
with blocklength $\ell^n$, rate $R$, 
and the block error probability at most $\ell^nR\epsilon$.  
Obviously, $P_\mathrm{e}(X\mid Y)$ is \statq\ of $(X, Y)$.
The average error probability $P_\mathrm{e}(X\mid Y)$ takes a value in $[0, (q-1)/q]$.
As it has been the case in the study of the binary case~\cite{korada2010polar}, 
the Bhattacharyya parameter is useful for bounding the error probability.
\begin{lemma}\label{lem:PeZ}
\begin{align*}
&\frac{q-1}{q^2}\left(\sqrt{1+(q-1)Z(X\mid Y)} - \sqrt{1-Z(X\mid Y)}\right)^2\\
&\quad\le P_\mathrm{e}(X\mid Y)\\
&\quad\le \min_{k=1,2,\dotsc,q-1}\left\{\frac{(q-1)Z(X\mid Y)+k(k-1)}{k(k+1)}\right\}.
\end{align*}
\end{lemma}
\noindent
Proof of Lemma~\ref{lem:PeZ} is in Appendix~\ref{sec:PeZ}.

Another quantity which we study in this section is 
the expected total variation distance $T(X\mid Y)$ between the a posteriori probability and the uniform distribution, 
defined as
\begin{equation*}
T(X\mid Y) := \sum_{y\in\mathcal{Y}} P_Y(y) \sum_{x\in\mathbb{F}_q} \left|P_{X\mid Y}(x\mid y) - \frac1q\right|.
\end{equation*}
Properties of the random quantity $T(\mathsf{X}_n\mid\mathsf{Y}_n)$ is 
important in polar codes for lossy source coding~\cite{korada2010lossy}, \cite{karzand2010polar}.
More precisely, if one has
\begin{equation*}
\Pr(T(\mathsf{X}_n\mid \mathsf{Y}_n)<\epsilon)\ge R
\end{equation*}
for the test channel $(\mathsf{X}_0, \mathsf{Y}_0)=(X,Y)$,
then there exists a polar code for source coding with blocklength $\ell^n (1-R)$, rate $1-R$ and
the average distortion at most $\mathcal{D} + d_{\max}\ell^n R\epsilon$
where $\mathcal{D}$ denotes the average distortion for the test channel and where $d_{\max}$ is the maximum value of
the distortion function~\cite{korada2010lossy}, \cite{karzand2010polar}.
Note that $T(X\mid Y)$ is \statq.
The total variation distance $T(X\mid Y)$ takes a value in $[0, 2(q-1)/q]$.
The following lemma establishes a relationship between 
the total variation distance $T(X\mid Y)$ and the average error 
probability $P_\mathrm{e}(X\mid Y)$.

\begin{lemma}\label{lem:TP}
\begin{align*}
&2\left(\frac{q-1}{q}-P_\mathrm{e}(X\mid Y)\right) \le T(X\mid Y)
\le 
\frac{2(q-1)}{q}\\
&\quad - \frac2q \max_{k=1,\dotsc,q-1} \bigg\{k(k+1)P_\mathrm{e}(X\mid Y) - k(k-1)\bigg\}.
\end{align*}
\end{lemma}
\noindent
The proof is in Appendix~\ref{sec:proofTP}.

The Fourier transform of the a posteriori probability is defined for analyzing $T(X\mid Y)$.
\begin{definition}[Character]
Let $\omega_p\in\mathbb{C}$ be a primitive complex $p$-th root of unity.
Define $\chi(x) := \omega_p^{\mathrm{Tr}(x)}$ for any $x\in\mathbb{F}_q$
where $\mathrm{Tr}: \mathbb{F}_q\to\mathbb{F}_p$ is defined as $x\mapsto \sum_{j=0}^{m-1} x^{p^j}$.
Here, $\mathrm{Tr}(x)\in\mathbb{F}_p$ appearing in the exponent 
should be regarded as an integer via the natural correspondence 
between $\mathbb{F}_p$ and $\mathbb{Z}/p\mathbb{Z}$.  
\end{definition}
From the definition of $\chi(x)$,
it satisfies the following properties.
\begin{align*}
\chi(0)&=1\\
|\chi(x)|&=1, \hspace{2em}\text{for any } x\in\mathbb{F}_q\\
\chi(x + z) &= \chi(x)\chi(z), \hspace{2em}\text{for any } x, z\in\mathbb{F}_q\\
\sum_{x\in\mathbb{F}_q}\chi(x)&=0.
\end{align*}
In this paper, we only use $\chi(x)$ through these properties.

\begin{definition}[Fourier transform]
For any fixed $y\in\mathcal{Y}$, the Fourier transform of the a posteriori probability $P_{X\mid Y}$ of a source $(X, Y)$ is defined as
\begin{equation*}
P_{X\mid Y}^*(w\mid y):= \sum_{z\in\mathbb{F}_q} P_{X\mid Y}(z\mid y) \chi(wz)
\end{equation*}
for $w\in\mathbb{F}_q$.
\end{definition}
Note that $P_{X\mid Y}^*(0\mid y) = 1$ for any $y\in\mathcal{Y}$.
Like the role of $Z(X\mid Y)$ in studying $P_\mathrm{e}(X\mid Y)$, 
the auxiliary quantity $S(X\mid Y)$, defined as 
\begin{align*}
S(X\mid Y) &:= \frac{1}{q-1}\sum_{w\in\mathbb{F}_q^\times} \sum_{y\in\mathcal{Y}}P_Y(y)\left|P_{X\mid Y}^{*}(w\mid y)\right|
\end{align*}
can be used for analyzing $T(X\mid Y)$.
The quantity $S(X\mid Y)$ takes a value in $[0,1]$.
Note that, although $S(X\mid Y)$ is identical to $T(X\mid Y)$ (and $1-2P_\mathrm{e}(X\mid Y)$) when $q=2$, 
$S(X\mid Y)$ is in general different from $T(X\mid Y)$.  
In this regard, consideration of the quantity $S(X\mid Y)$ 
is a novel idea that comes into play when one considers non-binary cases.  
Although $S(X\mid Y)$ is {\it not} invariant under arbitrary permutations of symbols in the a posteriori distribution,
$S(X\mid Y)$ is invariant under a permutation of symbols in the a posteriori distribution
when the permutation is addition or multiplication
on the finite field i.e., $S(X\mid Y) = S(r(Y) X+d(Y)\mid Y)$ for any $d:\mathcal{Y}\to\mathbb{F}_q$ and $r:\mathcal{Y}\to\mathbb{F}_q^\times$.
Hence, if $(X, Y)\overset{\mathrm{a}(1)}{\sim}(X', Y')$, it holds that $S(X\mid Y)=S(X'\mid Y')$.

The following lemma relates the quantity $S(X\mid Y)$ with 
the average error probability $P_\mathrm{e}(X\mid Y)$.  
\begin{lemma}\label{lem:SP}
\begin{align*}
& 1-\frac{q}{q-1}P_\mathrm{e}(X\mid Y)\le S(X\mid Y)\\
&\quad\le\min_{k=1,\dotsc,q-1}\Biggl\{
k(k+1)\\
&\qquad\cdot\Biggl[\left(\frac{k}{k+1}-P_\mathrm{e}(X\mid Y)\right)\sqrt{1- \frac{q}{q-1}\frac{k-1}k}\\
&\qquad+ \left(P_\mathrm{e}(X\mid Y)- \frac{k-1}{k}\right) \sqrt{1- \frac{q}{q-1}\frac{k}{k+1}}\Biggr]
\Biggr\}.
\end{align*}
\end{lemma}
\noindent
The proof is in Appendix~\ref{sec:proofSP}.

We now define the following equivalence relation 
for establishing relationship among several quantities 
for a source $(X, Y)$ defined so far.  
\begin{definition}
For $A(X\mid Y)\in[0,1]$ and $B(X\mid Y)\in[0,1]$, we say $A(X\mid Y) \overset{\mathrm{e}}{\sim} B(X\mid Y)$ if and only if
there exists $\epsilon>0$ and $c\in(0,1]$ such that
if $B(X\mid Y)<\epsilon$,
\begin{equation*}
 B(X\mid Y)^\frac1c \le A(X\mid Y) \le B(X\mid Y)^c
\end{equation*}
and if $1-B(X\mid Y)<\epsilon$,
\begin{equation*}
(1-B(X\mid Y))^\frac1c \le 1-A(X\mid Y) \le (1-B(X\mid Y))^c
\end{equation*}
for any source $(X, Y)$.
\end{definition}
From Lemmas~\ref{lem:PeZ},~\ref{lem:TP} and \ref{lem:SP}, the following corollary is obtained.
\begin{corollary}\label{cor:equiv}
$(q/(q-1))P_\mathrm{e}(X\mid Y)\overset{\mathrm{e}}{\sim} Z(X\mid Y)
\overset{\mathrm{e}}{\sim} 1-(q/(2(q-1)))T(X\mid Y) \overset{\mathrm{e}}{\sim} 1-S(X\mid Y)$.
\end{corollary}

The following four quantities are used in the derivation of the speed of polarization in the next subsection.
\begin{definition}\label{def:maxmin}
For any channel $(X, Y)$, $Z_{\max}(X, Y)$ and $Z_{\min}(X, Y)$ are defined as
\begin{align*}
Z_{\rm max}(X, Y) & := \max_{\substack{x\in\mathbb{F}_q, x'\in\mathbb{F}_q\\ x\ne x'}}
\sum_{y\in\mathcal{Y}} \sqrt{P_{Y\mid X}(y\mid x)P_{Y\mid X}(y\mid x')}\\
Z_{\rm min}(X, Y) & :=  \min_{x\in\mathbb{F}_q, x'\in\mathbb{F}_q}
\sum_{y\in\mathcal{Y}} \sqrt{P_{Y\mid X}(y\mid x)P_{Y\mid X}(y\mid x')}.
\end{align*}
For any source $(X, Y)$, $S_{\max}(X, Y)$ and $S_{\min}(X, Y)$ are defined as
\begin{align*}
S_{\max}(X, Y) &:= \max_{w\in\mathbb{F}_q^\times} \sum_{y\in\mathcal{Y}}P_Y(y)\left|P_{X\mid Y}^{*}(w\mid y)\right|\\
S_{\min}(X, Y) &:= \min_{w\in\mathbb{F}_q^\times} \sum_{y\in\mathcal{Y}}P_Y(y)\left|P_{X\mid Y}^{*}(w\mid y)\right|.
\end{align*}
\end{definition}
The quantities $Z_{\rm max}(X, Y)$ and $Z_{\rm min}(X, Y)$ are \statq.
Although $S_{\rm max}(X, Y)$ and $S_{\rm min}(X, Y)$ are {\it not} \statq,
it holds that $S_{\rm max / min}(X, Y) = S_{\rm max/ min}(r X+d(Y), Y)$ for any $d:\mathcal{Y}\to\mathbb{F}_q$ and $r\in\mathbb{F}_q^\times$.
Hence, if $(X, Y)\overset{\mathrm{a}(1)}{\sim}(X', Y')$,
it holds that $S_{\rm max/min}(X, Y)=S_{\rm max/min}(X', Y')$.
It is also straightforward to see the inequalities $Z_{\min}(X, Y)\le Z(X\mid Y)\le Z_{\max}(X, Y)$
and $S_{\min}(X, Y)\le S(X\mid Y)\le S_{\max}(X, Y)$ to hold.

\subsection{Speed of polarization}
In this subsection, we assume that $H(X\mid Y)\in (0,1)$, 
and also assume in view of Lemma~\ref{lem:s-c}, without loss of generality, 
that $(X, Y)$ is a channel.  
The exponents for channel coding and source coding are introduced in~\cite{korada2010polar}, \cite{korada2009thesis} for expressing the speed of polarization.
\begin{definition}
The exponent of $G$ for channel coding is defined as
\begin{equation*}
E_\mathrm{c}(G) := \frac1{\ell\log\ell}\sum_{i=0}^{\ell-1}\log D_\mathrm{c}^{(i)}(G)
\end{equation*}
where $D_\mathrm{c}^{(i)}(G)$ denotes the Hamming distance between the $i$-th row of $G$ and the linear space
spanned by $(i+1)$-th row to $(\ell-1)$-th row of $G$.
The exponent of $G$ for source coding is defined as
\begin{equation*}
E_\mathrm{s}(G) := \frac1{\ell\log\ell}\sum_{i=0}^{\ell-1}\log D_\mathrm{s}^{(i)}(G)
\end{equation*}
where $D_\mathrm{s}^{(i)}(G)$ denotes the Hamming distance between the $i$-th column of $G^{-1}$ and the linear space
spanned by $0$-th column to $(i-1)$-th column of $G^{-1}$.
\end{definition}
The following theorem holds, 
which was shown by Ar{\i}kan and Telatar~\cite{5205856}, Korada~et~al.~\cite{korada2010polar} and Korada~\cite{korada2009thesis} for the binary case 
with an additional condition.
\begin{theorem}\label{thm:speed}
If a channel $(X, Y)$ is polarized by $G$, it holds that for any $\epsilon > 0$,
\begin{equation} \label{eq:PolarPe}
\begin{split}
\lim_{n\to\infty} \Pr\left(P_\mathrm{e}(\mathsf{X}_n\mid \mathsf{Y}_n) < 2^{-\ell^{(E_\mathrm{c}(G)-\epsilon)n}}\right) &= 1-H(X\mid Y)\\
\lim_{n\to\infty} \Pr\left(P_\mathrm{e}(\mathsf{X}_n\mid \mathsf{Y}_n) < 2^{-\ell^{(E_\mathrm{c}(G)+\epsilon)n}}\right) &= 0.
\end{split}
\end{equation}
Furthermore, it holds that for any $\epsilon > 0$,
\begin{equation} \label{eq:PolarT}
\begin{split}
\lim_{n\to\infty} \Pr\left(T(\mathsf{X}_n\mid \mathsf{Y}_n) < 2^{-\ell^{(E_\mathrm{s}(G)-\epsilon)n}}\right) &= H(X\mid Y)\\
\lim_{n\to\infty} \Pr\left(T(\mathsf{X}_n\mid \mathsf{Y}_n) < 2^{-\ell^{(E_\mathrm{s}(G)+\epsilon)n}}\right) &= 0.
\end{split}
\end{equation}
\end{theorem}
\begin{remark}
Korada proved~\eqref{eq:PolarT} for the binary case with the aid of the condition $D_\mathrm{s}^{(i)}(G) \ge D_\mathrm{s}^{(i+1)}(G)$ for $i=0,\dotsc,\ell-2$~\cite{korada2009thesis}.
In this paper,~\eqref{eq:PolarT} is proved without any additional condition for both binary and non-binary cases.
\end{remark}
From Theorem~\ref{thm:speed}, the error probability of polar codes as channel codes of rate smaller than $I(W)$
and the distortion gap to the optimal distortion of polar codes as source codes are asymptotically bounded by $2^{-\ell^{(E_\mathrm{c}(G)-\epsilon)n}}$ and
$2^{-\ell^{(E_\mathrm{s}(G)-\epsilon)n}}$, respectively~\cite{korada2009thesis}.
From Corollary~\ref{cor:equiv}, it is sufficient to prove~\eqref{eq:PolarPe} and \eqref{eq:PolarT}
for $Z(\mathsf{X}_n\mid \mathsf{Y}_n)$ and $S(\mathsf{X}_n\mid \mathsf{Y}_n)$
instead of $P_\mathrm{e}(\mathsf{X}_n\mid \mathsf{Y}_n)$ and $T(\mathsf{X}_n\mid \mathsf{Y}_n)$, respectively.
The general proof shown in \cite{mori2010master, hassani2011rate} can be used for our purpose.
\begin{lemma}[\cite{mori2010master, hassani2011rate}]\label{lem:process}
Let $\{Z_n\}_{n=0,1,\dotsc}$ be a random process ranging in $[0,1]$
and $\{D_n\}_{n=0,1,\dotsc}$ be i.i.d.\ random variables ranging in $[1,\infty)$.
Assume that the expectation of $\log D_0$ exists.
Four conditions (c0)--(c3) are defined as follows.
\begin{itemize}
\item[(c0)] $Z_n\in(0,1]$ with probability 1.
\item[(c1)] There exists a random variable $Z_\infty$ such that $Z_n\to Z_\infty$ almost surely.
\item[(c2)] There exists a positive constant $c_0$ such that $Z_{n+1}\le c_0 Z_n^{D_n}$ with probability 1.
\item[(c3)] $Z_n^{D_n}\le Z_{n+1}$ with probability 1.
\end{itemize}
If (c0), (c1) and (c2) are satisfied, it holds that 
\begin{equation*}
\lim_{n\to\infty} \Pr\left(Z_n < 2^{-\ell^{(\mathbb{E}[\log_\ell D_0]-\epsilon)n}}\right) = \Pr(Z_\infty = 0).
\end{equation*}
If (c0), (c1) and (c3) are satisfied, it holds that 
\begin{equation*}
\lim_{n\to\infty} \Pr\left(Z_n < 2^{-\ell^{(\mathbb{E}[\log_\ell D_0]+\epsilon)n}}\right) = 0.
\end{equation*}
In the above, $\ell$ is any constant greater than 1.  
\end{lemma}
\begin{remark}
We do not assume the condition $Z_n<1$ to hold in Lemma~\ref{lem:process}, 
although it was assumed to hold with probability 1 in the arguments 
in~\cite{mori2010master} and \cite{hassani2011rate}. 
The condition is not needed in our argument here 
because we have only to deal with the case $Z_n\to 0$.
\end{remark}
From the assumption of Theorem~\ref{thm:speed}, the channel is polarized by $G$.
From Lemma~\ref{lem:HZ} and Corollary~\ref{cor:equiv}, $Z_{\max}(\mathsf{X}_n\mid \mathsf{Y}_n)$, $Z_{\min}(\mathsf{X}_n\mid \mathsf{Y}_n)$, $S_{\max}(\mathsf{X}_n\mid \mathsf{Y}_n)$ and $S_{\min}(\mathsf{X}_n\mid \mathsf{Y}_n)$ converge almost surely
to $\{0,1\}$-valued random variables.
From this observation, Lemma~\ref{lem:process} implies: 
\begin{itemize}
\item If the pair of $\{Z_n=Z_\mathrm{max}(\mathsf{X}_n, \mathsf{Y}_n)\}_{n=0,1,\dotsc}$ 
and $\{D_n=D_\mathrm{c}^{(B_n)}(G)\}_{n=0,1,\dotsc}$ satisfies (c0) and (c2), then 
the first equation of~\eqref{eq:PolarPe} holds.
\item If the pair of $\{Z_n=Z_\mathrm{min}(\mathsf{X}_n, \mathsf{Y}_n)\}_{n=0,1,\dotsc}$ 
and $\{D_n=D_\mathrm{c}^{(B_n)}(G)\}_{n=0,1,\dotsc}$ satisfies (c0) and (c3), then
the second equation of~\eqref{eq:PolarPe} holds.
\item If the pair of $\{Z_n=S_\mathrm{max}(\mathsf{X}_n, \mathsf{Y}_n)\}_{n=0,1,\dotsc}$ 
and $\{D_n=D_\mathrm{s}^{(B_n)}(G)\}_{n=0,1,\dotsc}$ satisfies (c0) and (c2), then 
the first equation of~\eqref{eq:PolarT} holds.
\item If the pair of $\{Z_n=S_\mathrm{min}(\mathsf{X}_n, \mathsf{Y}_n)\}_{n=0,1,\dotsc}$ 
and $\{D_n=D_\mathrm{s}^{(B_n)}(G)\}_{n=0,1,\dotsc}$ satisfies (c0) and (c3), then 
the second equation of~\eqref{eq:PolarT} holds.
\end{itemize}

The following lemma shows that the pair of 
$\{Z_n= Z_{\max}(\mathsf{X}_n, \mathsf{Y}_n)\}_{n=0,1,\dotsc}$ 
and $\{D_n = D_\mathrm{c}^{(B_n)}(G)\}_{n=0,1,\dotsc}$
satisfies the condition (c2),
and that the pair of 
$\{Z_n= Z_{\min}(\mathsf{X}_n, \mathsf{Y}_n)\}_{n=0,1,\dotsc}$ 
and $\{D_n = D_\mathrm{c}^{(B_n)}(G)\}_{n=0,1,\dotsc}$
satisfies the condition (c3).

\begin{lemma}[\cite{korada2010polar}]\label{lem:Zm}
For $i\in\{0,\dotsc,\ell-1\}$, it holds for any channel $(X, Y)$ that
\begin{align*}
Z_{\rm max}(X^{(i)}, Y^{(i)}) &\le q^{\ell-1-i} Z_{\rm max}(X, Y)^{D_\mathrm{c}^{(i)}(G)}\\
Z_{\rm min}(X, Y)^{D_\mathrm{c}^{(i)}(G)} &\le Z_{\rm min}(X^{(i)}, Y^{(i)}).
\end{align*}
\end{lemma}
The proof is omitted since the same proof for the binary alphabet in~\cite{korada2010polar} applies to the non-binary cases as well.
The following lemma shows that the pair of 
$\{Z_n= S_{\max}(\mathsf{X}_n, \mathsf{Y}_n)\}_{n=0,1,\dotsc}$ 
and $\{D_n = D_\mathrm{s}^{(B_n)}(G)\}_{n=0,1,\dotsc}$
satisfies the condition (c2),
and that the pair of 
$\{Z_n= S_{\min}(\mathsf{X}_n, \mathsf{Y}_n)\}_{n=0,1,\dotsc}$ 
and $\{D_n = D_\mathrm{s}^{(B_n)}(G)\}_{n=0,1,\dotsc}$
satisfies the condition (c3).
\begin{lemma}\label{lem:Sm}
For $i\in\{0,\dotsc,\ell-1\}$, it holds for any source $(X, Y)$ that
\begin{align*}
S_{\rm max}(X^{(i)}, Y^{(i)}) &\le q^{i} S_{\rm max}(X, Y)^{D_\mathrm{s}^{(i)}(G)}\\
S_{\rm min}(X, Y)^{D_\mathrm{s}^{(i)}(G)} &\le S_{\rm min}(X^{(i)}, Y^{(i)}).
\end{align*}
\end{lemma}
The proof is in Appendix~\ref{sec:proofS}.

Finally, we should prove that all the four processes satisfy (c0).
If the channel $(X,Y)$ satisfies the two inequalities 
$Z_{\min}(X,Y)>0$ and $S_{\min}(X,Y)>0$, 
(c0) obviously holds for the four processes since the property (c0)
is inherited in the processes i.e., if $Z_{\min}(X,Y)>0$, then $Z_{\min}(X^{(i)},Y^{(i)})>0$ for $i=0,\dotsc,\ell-1$.
In the following, we deal with the other cases.
It is sufficient to prove the following lemma.
\begin{lemma}\label{lem:ext}
Assume that $(X,Y)$ is polarized by $G$. Then,
\begin{align*}
\lim_{n\to\infty}\Pr\big(Z_{\rm min}(\mathsf{X}_n, \mathsf{Y}_n)>0\big)&=1\\
\lim_{n\to\infty}\Pr\big(S_{\rm min}(\mathsf{X}_n, \mathsf{Y}_n)>0\big)&=1.
\end{align*}
\end{lemma}
The proof is in Appendix~\ref{sec:proofext}.
Lemma~\ref{lem:ext} implies Theorem~\ref{thm:speed} for the cases $Z_\mathrm{min}(X,Y)=0$ or $S_\mathrm{min}(X, Y)=0$ due to the following reason.
For any $\delta>0$, there exists $n_0$ such that
\begin{align*}
\Pr\big(Z_{\rm min}(\mathsf{X}_n, \mathsf{Y}_n)>0\big)&\ge 1-\delta\\
\Pr\big(S_{\rm min}(\mathsf{X}_n, \mathsf{Y}_n)>0\big)&\ge 1-\delta
\end{align*}
for any $n\ge n_0$.
Theorem~\ref{thm:speed} can be applied 
to each of the channels $(X^{(b_1)\dotsm(b_{n_0})},Y^{(b_1)\dotsm(b_{n_0})})$ satisfying the inequalities $Z_\mathrm{min}(X^{(b_1)\dotsm(b_{n_0})},Y^{(b_1)\dotsm(b_{n_0})})>0$ and
$S_\mathrm{min}(X^{(b_1)\dotsm(b_{n_0})},Y^{(b_1)\dotsm(b_{n_0})})>0$.
As a consequence, it holds that for any $\delta>0$ and $\epsilon>0$
\begin{equation*}
\begin{split}
&\left(1-H(X\mid Y)\right)(1-\delta)\\
&\qquad\le \liminf_{n\to\infty} \Pr\left(P_\mathrm{e}(\mathsf{X}_n\mid \mathsf{Y}_n) < 2^{-\ell^{(E_\mathrm{c}(G)-\epsilon)n}}\right),\\
&\limsup_{n\to\infty}\Pr\left(P_\mathrm{e}(\mathsf{X}_n\mid \mathsf{Y}_n) < 2^{-\ell^{(E_\mathrm{c}(G)-\epsilon)n}}\right)\\
&\qquad\le \left(1-H(X\mid Y)\right)(1-\delta) + \delta,\\
&\limsup_{n\to\infty} \Pr\left(P_\mathrm{e}(\mathsf{X}_n\mid \mathsf{Y}_n) < 2^{-\ell^{(E_\mathrm{c}(G)+\epsilon)n}}\right) \le \delta.
\end{split}
\end{equation*}
Similar inequalities corresponding to~\eqref{eq:PolarT} also hold.
By letting $\delta\to 0$, Theorem~\ref{thm:speed} is obtained.

A more detailed asymptotic analysis depending on the rate can also be performed
as shown in~\cite{tanaka2010rre}, \cite{hassani2010scaling1}, \cite{mori2010master}, \cite{hassani2011rate} for the binary case.  
For example, under the condition that $G$ polarizes $(X, Y)$, 
one can prove that for $R\in(0, 1-H(X\mid Y))$, 
\begin{align*}
\lim_{n\to\infty}\Pr\biggl(&
P_\mathrm{e}(\mathsf{X}_n\mid\mathsf{Y}_n)\\
&\quad<2^{-\ell^{E_\mathrm{c}(G)n+\sqrt{V_\mathrm{c}(G)n}
Q^{-1}\left(\frac{R}{1-H(X\mid Y)}\right)+f(n)}}
\biggr)=R
\end{align*}
holds for an arbitrary function satisfying $f(n)=o(\sqrt{n})$, 
where 
\begin{equation*}
V_\mathrm{c}(G):=\frac{1}{\ell}\sum_{i=0}^{\ell-1}
(\log_{\ell} D_\mathrm{c}^{(i)}(G)-E_\mathrm{c}(G))^2
\end{equation*}
and where $Q^{-1}(\cdot)$ is the inverse function of the error function 
$Q(t):=\int_t^\infty e^{-z^2/2}\,dz/\sqrt{2\pi}$.  

In the binary case, any source is polarized by $G$ if and only if $E_\mathrm{c}(G)>0$~\cite{korada2010polar}.
The property also holds when $q$ is a prime
since the condition $E_\mathrm{c}(G)>0$ is equivalent to the condition that a standard form of $G$ is not the identity matrix. 
However, it no longer holds when $q$ is not a prime, 
in which case there may be sources which are not polarized by $G$ 
even if $E_\mathrm{c}(G)>0$, as shown in Section~\ref{subsec:necessity}. 
Since non-zero scalar multiplication of a column does not change the exponent $E_\mathrm{c}(G)$,
even if there are non-polarizing sources for $G$ satisfying $E_\mathrm{c}(G)>0$, 
appropriate scalar multiplication of a column of $G$
gives a matrix with the same exponent $E_\mathrm{c}(G)$ which polarizes any source. 

\section{Reed-Solomon matrix and its exponent}\label{sec:RS}
Let $\mathbb{F}_q=\{x_0,\dotsc,x_{q-1}\}$.
Let $a=[a_0,\dotsc,a_{k-1}]\in\mathbb{F}_q^k$ and $p_a(X)=a_0+a_1X+\dotsb+a_{k-1}X^{k-1}$.
The encoder of the $q$-ary extended Reed-Solomon code is defined as $\varphi(a):=[p_a(x_0),p_a(x_1),\dotsc,p_a(x_{q-1})]$.
Let $\alpha$ be a primitive element of $\mathbb{F}_q$.
When $x_{q-1}=0$ and $x_i=\alpha^{-i}$ for $i=0,\dotsc,q-2$,
the generator matrix of the $q$-ary extended Reed-Solomon code is a lower submatrix of the $q\times q$ matrix $G_{\rm RS}(q)$ over $\mathbb{F}_q$
which we call the Reed-Solomon matrix
\begin{equation*}
G_{\rm RS}(q):=
\begin{bmatrix}
1&1&1&\dotsm&1&0\\
1&\alpha&\alpha^2&\cdots&\alpha^{q-2}&0\\
1&\alpha^2&\alpha^4&\cdots&\alpha^{2(q-2)}&0\\
\vdots&\vdots&\vdots&\cdots&\vdots&\vdots\\
1&\alpha^{q-2}&\alpha^{2(q-2)}&\cdots&\alpha^{(q-2)(q-2)}&0\\
1&1&1&\cdots&1&1
\end{bmatrix}.
\end{equation*}
From Theorem~\ref{thm:main}, any source is polarized by the Reed-Solomon matrix.
Since extended Reed-Solomon codes are maximum distance separable (MDS) codes, 
one has $D_{\mathrm{c}}^{(i)}=i+1$ for $i=0,\dotsc,q-1$, and therefore 
the exponent of the Reed-Solomon matrix for channel coding is $E_\mathrm{c}(G_{\rm RS}(q))=\log(q!)/q$.
The inverse matrix of the Reed-Solomon matrix $G_{\rm RS}(q)$ is
\begin{equation*}
\begin{bmatrix}
1&1&1&\dotsm&1&0\\
1&\alpha^{-1}&\alpha^{-2}&\cdots&\alpha^{-(q-2)}&0\\
1&\alpha^{-2}&\alpha^{-4}&\cdots&\alpha^{-2(q-2)}&0\\
\vdots&\vdots&\vdots&\cdots&\vdots&\vdots\\
1&\alpha^{-(q-2)}&\alpha^{-2(q-2)}&\cdots&\alpha^{-(q-2)(q-2)}&0\\
1&0&0&\cdots&0&-1
\end{bmatrix}.
\end{equation*}
Hence, the exponent of the Reed-Solomon matrix for source coding is also $E_\mathrm{s}(G_{\rm RS}(q))=\log(q!)/q$.
Note that both of the exponents $\log(q!)/q$ monotonically increase in $q$ and converge to 1 as $q\to\infty$.

For $i\in\{0,1,\dotsc,q^n-1\}$, $i_ni_{n-1}\dotsb i_1$ denotes the $q$-ary expansion of $i$.
For polar codes constructed on the basis of the matrix $G_\mathrm{RS}(q)$, 
rows of $G_{\rm RS}(q)^{\otimes n}$ whose indices 
are in the set 
\begin{equation*}
\{i\in\{0,\dotsc,q^n-1\}\mid H(X^{(i_1)\dotsm(i_n)}\mid Y^{(i_1)\dotsm(i_n)}) < \epsilon\}
\end{equation*}
with some threshold $\epsilon>0$ are chosen, 
as mentioned in Section~\ref{sec:construction}.  
For the Reed-Muller codes, on the other hand, 
rows of $G_{\rm RS}(q)^{\otimes n}$ whose indices belong to 
\begin{equation*}
\{i\in\{0,\dotsc,q^n-1\}\mid i_1+ \dotsm+ i_n > n_0\}
\end{equation*}
are chosen 
for some threshold $n_0\in\{0,1,\dotsc,n(q-1)\}$%
\footnote{Here, $i_1,\dotsc,i_n$ are treated as integers in the additions.}.
In order to maximize the minimum distance, rows of $G_{\rm RS}(q)^{\otimes n}$ 
with indices in the set 
\begin{equation}\label{eq:hyp}
\{i\in\{0,\dotsc,q^n-1\}\mid (i_1+ 1)\dotsm(i_n+ 1) > n_0\}.
\end{equation}
with some threshold $n_0\in\{1,2,\dotsc,q^n\}$ should be chosen.
Hence, unless $q=2$, the selection rule for the Reed-Muller codes does not maximize the minimum distance.
Codes based on the selection rule \eqref{eq:hyp} are sometimes called Massey-Costello-Justesen codes~\cite{massey1973polynomial}
and hyperbolic cascaded Reed-Solomon codes~\cite{saints1993hyperbolic}.
Note that the minimum distance of Reed-Muller codes grows like $q^{n/2+o(n)}$
while the minimum distance of polar codes and hyperbolic codes grows like $q^{E_\mathrm{c}(G_{\rm RS}(q))n+o(n)}$.
From the above observation, the Reed-Solomon matrices can be regarded as a natural generalization of the matrix $\begin{bmatrix}1&0\\1&1\end{bmatrix}$ in the binary case.

We now consider the maximum exponent $E_{\max}(q, \ell):=\max_{G\in\mathbb{F}_q^{\ell\times\ell}}E_\mathrm{c}(G)$ 
for channel coding on given size $q$ of a finite field and size $\ell$ of a matrix.
For $q=2$, Korada et al.~\cite{korada2010polar} show that $E_{\max}(2, \ell)<0.55$ for $\ell\le 31$,
and also show a method of construction of binary matrices with large exponents using the Bose-Chaudhuri-Hocquenghem (BCH) codes.
For $q\ge2$ and $\ell\le q$, the $\ell\times\ell$ lower-right submatrix of the $q$-ary Reed-Solomon matrix gives the largest exponent
so that $E_{\max}(q,\ell)=\log (\ell!)/(\ell\log\ell)$ for $\ell\le q$ since the Reed-Solomon code is an MDS code~\cite{macwilliams1988tec}.
Thus, the Reed-Solomon matrices with $q>2$ can be regarded as providing 
a systematic means to construct polar codes with larger exponents for the case $\ell\le q$. 
For example, for $q=4$, $E_{\max}(4, 4) = E_\mathrm{c}(G_{\rm RS}(4)) \approx 0.573\,12$,
which is larger than the upper bound $0.55$ of $E_{\max}(2, \ell)$ for $\ell\le 31$.
For $\ell > q > 2$, on the other hand, 
algebraic geometry codes are considered to be useful since they have a large minimum distance and the nested structure which are plausible
in making $D_\mathrm{c}^{(i)}$s larger.
The examples using the Hermitian codes are shown in~\cite{mori2010non}, 
in which $q=p^{m}$ and $\ell=p^{3m/2}$ for an even integer $m$.  
The $q$-ary $\ell\times\ell$ matrix constructed
on the basis of the Hermitian code has a yet larger exponent 
than the Reed-Solomon matrix $G_{\rm RS}(q)$ for $q>4$.

\section{Numerical results}\label{sec:sim}
\begin{figure}[tb]
\begin{center}
\includegraphics[width=\hsize]{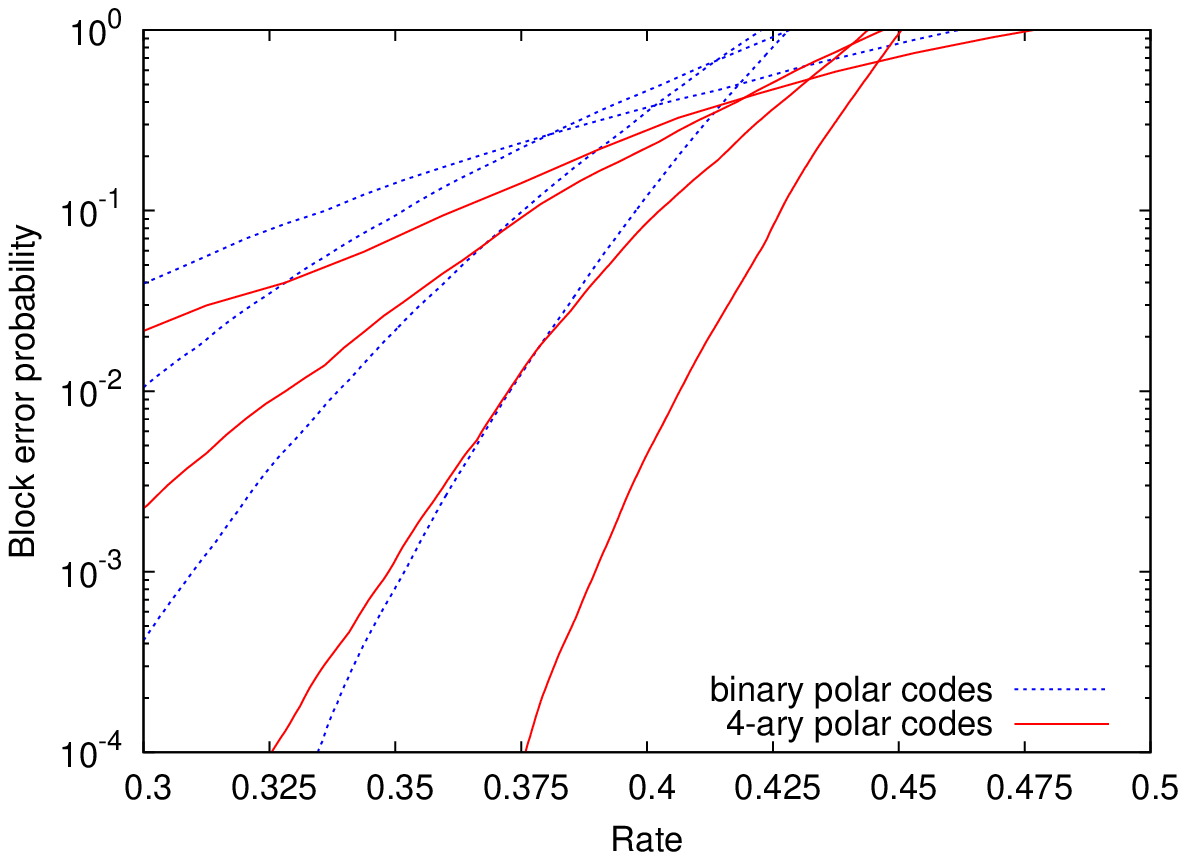}
\caption{
Numerical results on the upper bound of the block error probability of 
polar codes over an AWGN channel, for which the standard deviation of noise is set equal to 0.978\,65\@.
The capacity of the AWGN channel is about 0.5\@.
Results of binary polar codes and quaternary polar codes using $G_{\rm RS}(4)$
are shown by dotted curves and solid curves, respectively.
Blocklengths are $2^7$, $2^9$, $2^{11}$, and $2^{13}$ viewed as binary codes.
}
\label{fig:RS_awgn}
\end{center}
\end{figure}

In Fig.~\ref{fig:RS_awgn}, performance of 
the original binary polar codes with $G_1$ and quaternary polar codes using the Reed-Solomon matrix $G_{\rm RS}(4)$ are compared on the binary-input additive-white-Gaussian-noise (AWGN) channel with capacity about 0.5.
Instead of the actual error probability, the upper bound $\sum_{i\in \mathcal{A}}P_\mathrm{e}(X^{(i_1)\dotsm(i_n)}\mid Y^{(i_1)\dotsm(i_n)})$ is plotted
where $\mathcal{A}$ denotes the set of chosen row indices 
in constructing polar codes.
This bound is accurate for rates not close to the capacity~\cite{5205857}.
A significant improvement by the quaternary polar codes 
over the binary counterparts is observed in terms of the block error probability, although the error probability of the quaternary polar codes is still larger than that of (3,6)-regular low-density-parity-check (LDPC) codes
except in a low-rate region.

\section{Summary}\label{sec:sum}
We have shown that 
a necessary and sufficient condition for a $q$-ary $\ell\times\ell$ invertible matrix $G$ over $\mathbb{F}_q$ with a non-identity
standard form $\bar{G}$ to polarize any source/channel 
is $\mathbb{F}_p(\bar{G})=\mathbb{F}_q$.
The result about speed of polarization for the binary alphabet has been generalized to non-binary cases.
We have also explicitly given $q$-ary $\ell\times\ell$ matrices 
with $\ell\le q$ 
on the basis of the $q$-ary Reed-Solomon matrices, which have 
the largest exponent $E_\mathrm{max}(q, \ell)=\log(\ell!)/(\ell\log\ell)$ 
among all $\ell\times\ell$ matrices.  
Performance of non-binary polar codes based on Reed-Solomon matrices 
are found via numerical evaluation to be significantly better than the performance of the original binary polar codes.

\appendices

\section{Proof of Lemma~\ref{lem:Zd}}\label{sec:proofZd}
In order to relate the entropy and the Bhattacharyya parameter, the following lemma is useful.
\begin{lemma}[{\cite[Sec.\ 5.6]{gallager1968information}}]\label{lem:cutoff}
For any random variables $X$, $Y$ and $Z$ on sets $\mathcal{X}$, $\mathcal{Y}$ and $\mathcal{Z}$, respectively,
\begin{align*}
&\sum_{x\in\mathcal{X}, y\in\mathcal{Y}} P_{X, Y}(x, y) \log\frac{P_{X,Y}(x,y)}{P_X(x)P_Y(y)}\\
&\quad\ge -\log \sum_{y\in\mathcal{Y}} \left(\sum_{x\in\mathcal{X}} P_X(x)\sqrt{P_{Y\mid X}(y\mid x)}\right)^2\\
&\sum_{x\in\mathcal{X}, y\in\mathcal{Y}, z\in\mathcal{Z}} P_{X, Y, Z}(x, y, z)
\log\frac{P_{X,Y\mid Z}(x,y\mid z)}{P_{X\mid Z}(x\mid z)P_{Y\mid Z}(y\mid z)}\\
&\quad\ge -\log \sum_{y\in\mathcal{Y}, z\in\mathcal{Z}}P_Z(z)\\
&\quad\cdot\left(\sum_{x\in\mathcal{X}} P_{X\mid Z}(x\mid z)\sqrt{P_{Y\mid X, Z}(y\mid x, z)}\right)^2.
\end{align*}
\end{lemma}
The second inequality is an immediate consequence of the first inequality and Jensen's inequality.
The first inequality is obtained in \cite[Sec.\ 5.6]{gallager1968information}.
In Lemma~\ref{lem:cutoff}, the quantities on the left-hand sides are 
the mutual information between $X$ and $Y$, and the conditional mutual information between $X$ and $Y$ given $Z$, respectively.
The quantities on the right-hand sides are the cutoff rate and the conditional cutoff rate, respectively.

Given a source $(X, Y)$, 
let $(U_0,U_1,X_0,X_1,Y_0,Y_1)$ be the random variables 
defined by applying the basic transform with $G_\gamma$ to the source $(X, Y)$, 
as described in Section~\ref{sec:polar}.
Then, one obtains
\begin{align*}
&H(U_1\mid Y_1) - H(U_1\mid U_0, Y_0,Y_1)\\
&= \sum_{u_0\in\mathbb{F}_q,u_1\in\mathbb{F}_q,y_0\in\mathcal{Y},y_1\in\mathcal{Y}} P_{U_0,U_1,Y_0,Y_1}(u_0,u_1,y_0,y_1)\\
&\quad\cdot\log\frac{P_{U_0,U_1,Y_0\mid Y_1}(u_0,u_1,y_0\mid y_1)}{P_{U_1\mid Y_1}(u_1\mid y_1)P_{U_0,Y_0\mid Y_1}(u_0,y_0\mid y_1)}\\
&\ge -\log \Biggl[\sum_{y_1\in\mathcal{Y}} P_Y(y_1) \sum_{u_0\in\mathbb{F}_q, y_0\in\mathcal{Y}}\\
&\quad\Biggl[\sum_{u_1\in\mathbb{F}_q} P_{U_1\mid Y_1}(u_1\mid y_1)\sqrt{P_{U_0,Y_0\mid U_1, Y_1}(u_0,y_0\mid u_1, y_1)}\Biggr]^2\Biggr]\\
&= -\log \Biggl[\sum_{y_1\in\mathcal{Y}} P_Y(y_1)
\sum_{u_0\in\mathbb{F}_q, y_0\in\mathcal{Y}}\\
&\quad\Biggl[\sum_{u_1\in\mathbb{F}_q} P_{U_1\mid Y_1}(u_1\mid y_1)\sqrt{P_{X_0,Y_0}(u_0+u_1,y_0)}\Biggr]^2\Biggr]\\
&= -\log \Biggl[\sum_{y_1\in\mathcal{Y}} P_Y(y_1)\\
&\quad\cdot\sum_{u_1\in\mathbb{F}_q, u_1'\in\mathbb{F}_q}
P_{X\mid Y}(\gamma u_1\mid y_1)P_{X\mid Y}(\gamma u_1'\mid y_1)\\
&\quad\cdot\sum_{u_0\in\mathbb{F}_q, y_0\in\mathcal{Y}}\sqrt{P_{X,Y}(u_0+u_1,y_0)}\sqrt{P_{X,Y}(u_0+u_1',y_0)}\Biggr]\\
&= -\log \Biggl[1-\sum_{y_1\in\mathcal{Y}} P_Y(y_1)\\
&\quad\cdot\sum_{u_1\in\mathbb{F}_q, u_1'\in\mathbb{F}_q}
P_{X\mid Y}(\gamma u_1\mid y_1)P_{X\mid Y}(\gamma u_1'\mid y_1)
\Biggl(1\\
&\,-\sum_{u_0\in\mathbb{F}_q, y_0\in\mathcal{Y}}\sqrt{P_{X,Y}(u_0+u_1,y_0)}\sqrt{P_{X,Y}(u_0+u_1',y_0)}\Biggr)\Biggr]\\
&= -\log \Biggl[1-q\sum_{d\in\mathbb{F}_q}\sum_{y_1\in\mathcal{Y}} P_Y(y_1)\\
&\quad\cdot\sum_{u_1\in\mathbb{F}_q}\frac1q
P_{X\mid Y}(\gamma u_1\mid y_1)P_{X\mid Y}(\gamma u_1 + \gamma d\mid y_1)
\Biggl(1\\
&\quad-\sum_{u_0\in\mathbb{F}_q, y_0\in\mathcal{Y}}\sqrt{P_{X,Y}(u_0+u_1,y_0)}\\
&\quad\cdot\sqrt{P_{X,Y}(u_0+u_1+d,y_0)}\Biggr)\Biggr]\\
&\ge -\log \Biggl[1-
q\sum_{d\in\mathbb{F}_q^\times}
\Biggl(\sum_{u_1\in\mathbb{F}_q, y_1\in\mathcal{Y}} \frac1q P_Y(y_1) \\
&\quad\cdot\sqrt{P_{X\mid Y}(\gamma u_1\mid y_1)P_{X\mid Y}(\gamma u_1 + \gamma d\mid y_1)}\Biggr)^2\\
&\quad\cdot \Biggl(1-\sum_{u_0\in\mathbb{F}_q, y_0\in\mathcal{Y}}\sqrt{P_{X,Y}(u_0,y_0)}\sqrt{P_{X,Y}(u_0+d,y_0)}\Biggr)\Biggr]\\
&= -\log \Bigg[1-
\frac1q\sum_{d\in\mathbb{F}_q^\times}
Z_{\gamma d}(X\mid Y)^2 (1-Z_d(X\mid Y))\Bigg].
\end{align*}
The first and second inequalities are obtained by Lemma~\ref{lem:cutoff} and Jensen's inequality, respectively.

The assumption of Lemma~\ref{lem:Zd} implies that the above formula 
evaluated for $(X,Y)=(X_{(n)},Y_{(n)})$ approaches 0 as $n\to\infty$, 
or equivalently, that for any $\epsilon>0$, there exists $n_0$ such that
\begin{equation*}
Z_{\gamma d}(X_{(n)}\mid Y_{(n)}) (1-Z_d(X_{(n)}\mid Y_{(n)})) < \epsilon
\end{equation*}
for any $n\ge n_0$ and any $d\in\mathbb{F}_q^\times$.
Fix $\epsilon\in(0,1/2)$.
Then, there exists $n_0$ such that
\begin{equation*}
Z_{\gamma d}(X_{(n)}\mid Y_{(n)}) (1-Z_d(X_{(n)}\mid Y_{(n)})) < \epsilon^2
\end{equation*}
for any $n\ge n_0$ and any $d\in\mathbb{F}_q^\times$, 
which in turn implies 
\begin{equation*}
Z_{\gamma d}(X_{(n)}\mid Y_{(n)}) < \epsilon\hspace{2em} \text{ or } \hspace{2em}
1-Z_{d}(X_{(n)}\mid Y_{(n)}) < \epsilon
\end{equation*}
for any $n\ge n_0$ and any $d\in\mathbb{F}_q^\times$.
Assume
$1-Z_{d'}(X_{(n')}\mid Y_{(n')}) < \epsilon$ for fixed $n'\ge n_0$ and fixed $d'\in\mathbb{F}_q^\times$.
Then, from
\begin{equation*}
Z_{d'}(X_{(n')}\mid Y_{(n')}) (1-Z_{\gamma^{-1} d'}(X_{(n')}\mid Y_{(n')})) < \epsilon^2
\end{equation*}
one obtains $1-Z_{\gamma^{-1} d'}(X_{(n')}\mid Y_{(n')})) < \epsilon^2/(1-\epsilon) < \epsilon$.
By iterating this procedure, one proves that 
$1-Z_{\gamma^{i} d'}(X_{(n')}\mid Y_{(n')})) < \epsilon$ holds for all $i\in\{0,\dotsc,q-2\}$.
In the same way, 
when $Z_{d'}(X_{(n')}\mid Y_{(n')}) < \epsilon$ is assumed for fixed $n'\ge n_0$ and fixed $d'\in\mathbb{F}_q^\times$,
one can prove that $Z_{\gamma^{i} d'}(X_{(n')}\mid Y_{(n')})) < \epsilon$ holds for all $i\in\{0,\dotsc,q-2\}$.
This completes the proof of Lemma~\ref{lem:Zd}. 

\section{Bhattacharyya parameter and error probability}\label{sec:PeZ}
\begin{figure}[t]
\begin{center}
\psfrag{UBound}{\tiny Lemma~\ref{lem:PZU}}
\psfrag{LBound}{\tiny Lemma~\ref{lem:PZL}}
\psfrag{CUBound}{\tiny Eq.~\eqref{eq:PZCU}}
\psfrag{Z}{$Z(X\mid Y)$}
\psfrag{P}{$P_\mathrm{e}(X\mid Y)$}
\includegraphics[height=0.25\vsize]{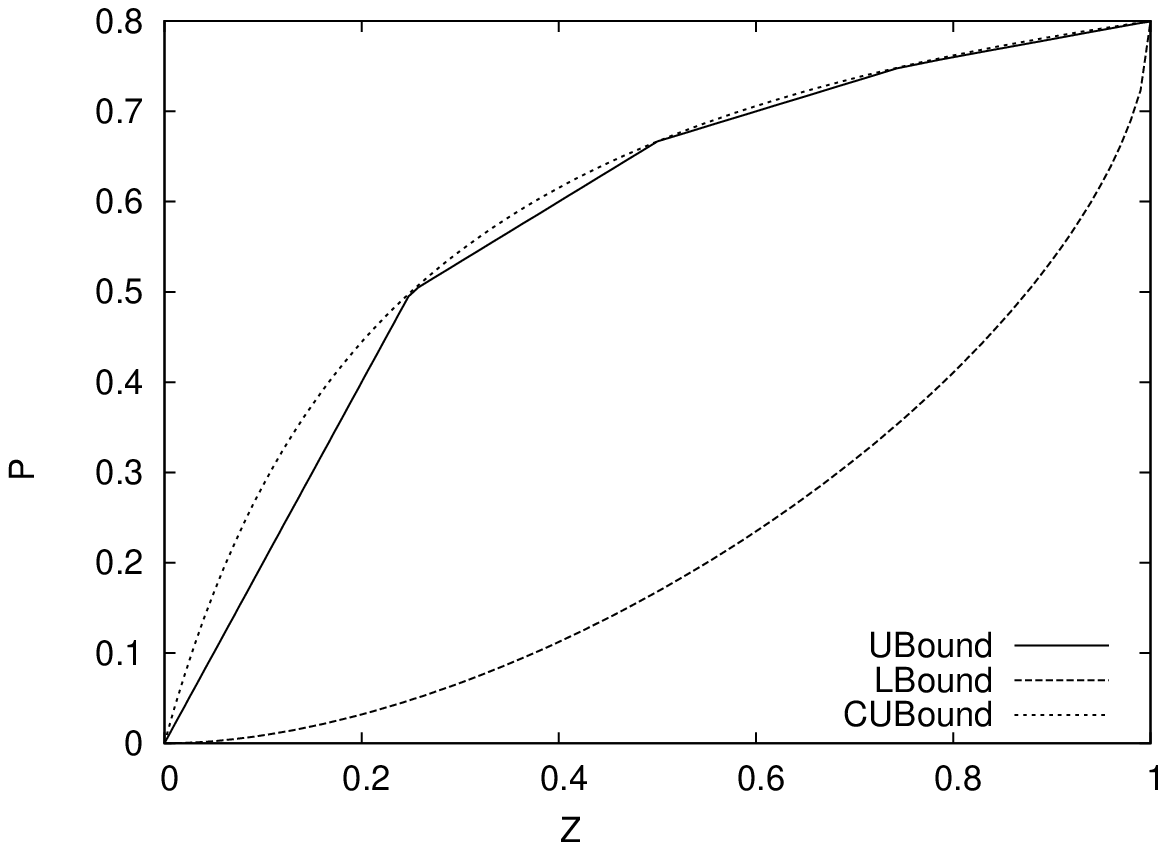}
\caption{
The upper and lower bounds of error probability for $q=5$.
}
\label{fig:PeZ}
\end{center}
\end{figure}

In this appendix, an unconditional version of 
Lemma~\ref{lem:PeZ} is proved.
Lemma~\ref{lem:PeZ} itself is then proved straightforwardly by Jensen's inequality.
For the proof of the unconditional version, 
one can regard $\mathcal{X}$ as any finite set whose size $q$ is not necessarily a power of a prime.
Let $X$ be a random variable on $\mathcal{X}$. 
The optimum estimator for $X$ minimizing the probability of error 
is given by $\hat{x}:=\arg\max_xP_X(x)$, 
with the error probability 
\begin{equation*}
P_\mathrm{e}(X):=1-\max_{x\in\mathcal{X}}P_X(x).
\end{equation*}
The Bhattacharyya parameter is defined as 
\begin{equation*}
Z(X):=\frac{1}{q-1}\sum_{\substack{x\in\mathcal{X}, x'\in\mathcal{X},\\ x'\not=x}}\sqrt{P_X(x)P_X(x')}.
\end{equation*}
The following lemma gives an upper bound of the error probability 
in terms of the Bhattacharyya parameter. 
\begin{lemma}\label{lem:PZU}
\begin{equation*}
P_\mathrm{e}(X) \le \min_{k=1,2,\dotsc,q-1}\left\{\frac{(q-1)Z(X)+k(k-1)}{k(k+1)}\right\}.
\end{equation*}
\end{lemma}
\begin{IEEEproof}
Noting that $P_X(\hat{x})=1-P_\mathrm{e}(X)$ holds by the definition, one has 
\begin{equation*}
\sum_{x}\sqrt{P_X(x)} = \sqrt{1-P_\mathrm{e}(X)} + \sum_{x\ne \hat{x}}\sqrt{P_X(x)}.
\end{equation*}
In order to prove the lemma, we first find the extremal distribution 
of $X$ for which $Z(X)$ is minimized with $P_\mathrm{e}(X)$ fixed. 
As we will show, this amounts to minimizing the second term on the right-hand side with respect to $P_X(x)$ 
under the constraint that the error probability is $P_\mathrm{e}(X)$. 
We thus consider the following minimization problem for $\{p_i\}_{i=0,1,\dotsc,q-2}$.
\begin{align*}
\text{minimize:}\hspace{2em}& \sum_i\sqrt{p_i}\\
\text{subject to:}\hspace{2em}& \sum_i p_i = P_\mathrm{e}(X)\\
& 0\le p_i \le 1-P_\mathrm{e}(X).
\end{align*}
Let $\{p_i^*\}_{i=0,1,\dotsc,q-2}$ be the optimum solution of the minimization problem.
Since $\sqrt{x}$ is a concave function, $p_i^*$ is 0 or $1-P_\mathrm{e}(X)$ except for at most one $i$~\cite{feder1994relations}.
Let $t-1$ be the number of $p_i^*$s which are equal to $1-P_\mathrm{e}(X)$, 
then $t=\lfloor1/(1-P_\mathrm{e}(X))\rfloor$ holds. 
The value of $p_i^*$ which is not 0 or $1-P_\mathrm{e}(X)$ 
is equal to $1-t(1-P_\mathrm{e}(X))$.
Hence,
\begin{equation}
\sum_{x}\sqrt{P_X(x)} \ge t\sqrt{1-P_\mathrm{e}(X)} + \sqrt{1-t(1-P_\mathrm{e}(X))}.\label{eq:UZP}
\end{equation}
By squaring both sides of \eqref{eq:UZP}, one obtains the inequality 
\begin{align*}
1+(q-1)Z(X) &\ge 1 + t(t-1)(1-P_\mathrm{e}(X))\\
&\quad + 2t\sqrt{(1-P_\mathrm{e}(X))(1-t(1-P_\mathrm{e}(X))}
\end{align*}
which implies the minimum achievable value of the Bhattacharyya parameter 
for a given error probability. 
The right-hand side of the above inequality is further lower bounded 
by applying the inequality $1-P_\mathrm{e}(X)\ge 1-t(1-P_\mathrm{e}(X)) \Leftrightarrow t\ge 1/(1-P_\mathrm{e}(X))-1$ 
to the last term, yielding 
\begin{align}
(q-1)Z(X) &\ge t(t-1)(1-P_\mathrm{e}(X)) + 2t(1-t(1-P_\mathrm{e}(X)))\nonumber\\
&= -(1-P_\mathrm{e}(X))t^2+(1+P_\mathrm{e}(X))t.
\label{eq:UZP1}
\end{align}
Since the quadratic function $-(1-P_\mathrm{e}(X))x^2 + (1+P_\mathrm{e}(X))x$ is concave and 
takes a maximum at $x=(1+P_\mathrm{e}(X))/(2(1-P_\mathrm{e}(X))$, which is the center of the unit interval $[P_\mathrm{e}(X)/(1-P_\mathrm{e}(X)), 1/(1-P_\mathrm{e}(X))]$ containing $t$,
the inequality \eqref{eq:UZP1} still holds 
even if $t$ is replaced by any integer $k=1,2,\dotsc,q-1$.
\end{IEEEproof}
By replacing $t$ by $1/(1-P_\mathrm{e}(X))$ in \eqref{eq:UZP1}, one obtains a looser but smooth bound
\begin{equation}\label{eq:PZCU}
P_\mathrm{e}(X) \le \frac{(q-1)Z(X)}{(q-1)Z(X)+1}.
\end{equation}
This bound is also obtained from the monotonicity of the R\'{e}nyi entropy i.e., $H_{1/2}(X) \ge H_\infty(X)$ 
where $H_\alpha(X):=(1-\alpha)^{-1}\log\sum_xP_X(x)^\alpha$.
These upper bounds are plotted in Fig.~\ref{fig:PeZ} for $q=5$.

The next lemma provides a lower bound of the error probability 
in terms of the Bhattacharyya parameter. 
\begin{lemma}\label{lem:PZL}
\begin{equation*}
P_\mathrm{e}(X) \ge \frac{q-1}{q^2}\left(\sqrt{1+(q-1)Z(X)} - \sqrt{1-Z(X)}\right)^2.
\end{equation*}
\end{lemma}
\begin{IEEEproof}
We start with the same formula as that used as the starting point 
of the proof of Lemma~\ref{lem:PZU}. 
\begin{align}
\sum_{x}\sqrt{P_X(x)} &= \sqrt{1-P_\mathrm{e}(X)} + \sum_{x\ne \hat{x}}\sqrt{P_X(x)}\nonumber\\
&= \sqrt{1-P_\mathrm{e}(X)} + (q-1)\sum_{x\ne \hat{x}}\frac1{q-1}\sqrt{P_X(x)}\nonumber\\
&\le \sqrt{1-P_\mathrm{e}(X)} + (q-1)\sqrt{\frac1{q-1}P_\mathrm{e}(X)}\nonumber\\
&= \sqrt{1-P_\mathrm{e}(X)} + \sqrt{(q-1)P_\mathrm{e}(X)}.\label{eq:l1max}
\end{align}
The above inequality is obtained from Jensen's inequality.
This proof is the same as the proof of Fano's inequality for the R\'{e}nyi entropy~\cite{ben1978renyi}.
By squaring both sides of the above inequality, one has 
\begin{align*}
&1+(q-1)Z(X) \le \left[\sqrt{1-P_\mathrm{e}(X)}+\sqrt{q-1}\sqrt{P_\mathrm{e}(X)}\right]^2\\
&\Longleftrightarrow Z(X)\\
&\qquad \le \frac{(q-2) P_\mathrm{e}(X) + 2\sqrt{q-1}\sqrt{P_\mathrm{e}(X)(1-P_\mathrm{e}(X))}}{q-1}.
\end{align*}
The function 
\begin{equation*}
f(x) := \frac{(q-2)x + 2\sqrt{q-1}\sqrt{x(1-x)}}{q-1}
\end{equation*}
defined for $x\in[0,(q-1)/q]$
is continuous and strictly increasing
since
\begin{align*}
f'(x) &= \frac{q-2}{q-1} + \frac{1-2x}{\sqrt{q-1}\sqrt{x(1-x)}}\\
f''(x) &= -\frac1{2\sqrt{q-1}(x(1-x))^{3/2}} < 0
\end{align*}
and $f'((q-1)/q)=0$.
Hence, $f^{-1}(Z(X))\le P_\mathrm{e}(X)$ where the inverse function $f^{-1}(x)$ of $f(x)$ is
\begin{equation*}
f^{-1}(x) = \frac{q-1}{q^2}\left(\sqrt{1+(q-1)x} - \sqrt{1-x}\right)^2.
\end{equation*}
\end{IEEEproof}

Lemma~\ref{lem:PeZ} is obtained from Lemmas~\ref{lem:PZU} and \ref{lem:PZL} 
by applying Jensen's inequality. 
The lower and upper bounds are plotted in Fig.~\ref{fig:PeZ} for $q=5$.
The bounds given in Lemma~\ref{lem:PeZ} are the tightest among those which are given in terms of the Bhattacharyya parameter only.
Tight examples are shown below.
The lower bound in Lemma~\ref{lem:PeZ} is tight for the $q$-ary symmetric channel, defined by $\mathcal{X}=\mathcal{Y}=\{0,\dotsc,q-1\}$ and
\begin{equation*}
P_{Y\mid X}(y\mid x) = \begin{cases}
1-\epsilon,& \text{if } y = x\\
\epsilon/(q-1),& \text{if } y \ne x
\end{cases}
\end{equation*}
for $\epsilon\in[0,(q-1)/q]$.
In this case,
\begin{align*}
P_\mathrm{e}(X\mid Y) &= \epsilon\\
Z(X\mid Y) &= \frac{q-2}{q-1}\epsilon + 2\sqrt{\frac{\epsilon(1-\epsilon)}{q-1}}
\end{align*}
which satisfies the lower bound with equality.
The upper bound in Lemma~\ref{lem:PeZ} is tight for the following channel.
Let $\mathcal{X} = \{0,\dotsc,q-1\}$.
For fixed $k\in\{1,\dotsc,q-1\}$, 
let $\mathcal{Y} = {A_k\cup A_{k+1}}$
where $A_k:=\{\mathcal{A}\subseteq\mathcal{X}\mid |\mathcal{A}| = k\}$, and let 
\begin{equation*}
P_{Y\mid X}(y\mid x) = \begin{cases}
(1-\epsilon)/\binom{q-1}{k-1},& \text{if $|y|=k$ and } x\in y\\
\epsilon/\binom{q-1}{k},& \text{if $|y|=k+1$ and } x\in y\\
0,& \text{otherwise}
\end{cases}
\end{equation*}
for $\epsilon\in[0,1]$.
That is, the output of the channel is a subset of $\mathcal{X}$ 
containing the input $x$ and with size $k$ or $k+1$.  
This channel satisfies the upper bound with equality since it holds 
\begin{align*}
P_\mathrm{e}(X\mid Y) &= \frac{k^2-1+\epsilon}{k(k+1)}\\
Z(X\mid Y) &= \frac{k-1+\epsilon}{q-1}.
\end{align*}

\section{Proof of Lemma~\ref{lem:TP}}\label{sec:proofTP}
Similarly to Appendix~\ref{sec:PeZ}, it is sufficient to prove an unconditional version of the inequalities in Lemma~\ref{lem:TP}.  
Let $\mathcal{X}$ be a finite set of size $q$, 
and let $X$ be a random variable on $\mathcal{X}$.  
Let
\begin{equation*}
T(X) := \sum_{x\in\mathcal{X}} \left|P_X(x)-\frac1q\right|
\end{equation*}
be the total variation distance between $P_X$ and 
the uniform distribution over $\mathcal{X}$.

Let $t := \lfloor 1/(1-P_\mathrm{e}(X))\rfloor$.
The same argument as that of minimizing the concave function 
in Appendix~\ref{sec:PeZ} applies to minimizing $-T(X)$ given $P_\mathrm{e}(X)$, 
yielding the upper bound 
\begin{align*}
&T(X)\le t\Big(1-P_\mathrm{e}(X)-\frac1q\Big) + \left|1-t(1-P_\mathrm{e}(X))-\frac1q\right|\\
&\quad + (q-t-1)\frac1q\\
&= \frac{q-1}q + t \Big(1-P_\mathrm{e}(X)-\frac2q\Big) + \left|\frac{q-1}q - t(1-P_\mathrm{e}(X))\right|\\
&=: f_T(P_\mathrm{e}(X)).
\end{align*}
We now derive the concave hull of $f_T(x)$ for obtaining the upper bound of $T(X\mid Y)$.
Let $k$ be a positive integer smaller than $q$.
When $x$ satisfies $(k-1)/k\le x < k/(k+1)$, 
one has $k\le 1/(1-x)< k+1$, so that 
the value of $t=\lfloor 1/(1-x)\rfloor$ is 
equal to the constant $k$.  
The function $f_T(x)$ is hence 
a convex function of $x$ in the interval $(k-1)/k\le x < k/(k+1)$, and the linear interpolation of the values of $f_T(x)$ 
at the two endpoints $x=(k-1)/k$ and $x\uparrow k/(k+1)$ 
thus gives the concave hull of $f_T(x)$ for $(k-1)/k\le x<k/(k+1)$.  
One therefore obtains the inequality 
\begin{align*}
f_T(x)
&\le
(k+1)k\biggl[\bigl(k/(k+1)-x\bigr) f_T((k-1)/k)\\
&\quad + \bigl(x-(k-1)/k\bigr) \lim_{x\uparrow k/(k+1)}f_T(x)\biggr]
\end{align*}
for $x$ satisfying $(k-1)/k\le x < k/(k+1)$.  
By substituting
\begin{align*}
f_T((k-1)/k) &= \frac2q (q - k)\\
\lim_{x\uparrow k/(k+1)} f_T(x)
&=  \frac2q (q-k-1)
\end{align*}
one obtains
\begin{align}
f_T(x)
&\le
\frac2q(q-k-1)
+ \frac2q k
\left[\big(k-(k+1)x\big)\right]\nonumber\\
&=
\frac{2(q-1)}q - \frac2q[-(1-x)k^2+(1+x)k]
\nonumber
\end{align}
and therefore 
\begin{equation}
T(X)\le
\frac{2(q-1)}q - \frac2q[-(1-P_\mathrm{e}(X))k^2+(1+P_\mathrm{e}(X))k]
\label{eq:TPU}
\end{equation}
for $P_\mathrm{e}(X)$ satisfying $(k-1)/k\le P_\mathrm{e}(X)<k/(k+1)$.  
As shown in the proof of Lemma~\ref{lem:PZU},
the inequality~\eqref{eq:TPU} is correct for any $P_\mathrm{e}(X)\in[0,(q-1)/q]$.
Note that by replacing $k$ by $1/(1-P_\mathrm{e}(X))$, one obtains a looser but smooth upper bound
\begin{equation*}
f_T(P_\mathrm{e}(X))\le \frac2{1-P_\mathrm{e}(X)}\left(\frac{q-1}q - P_\mathrm{e}(X)\right).
\end{equation*}

The unconditional version of the other inequality in Lemma~\ref{lem:TP} is obtained 
by applying the triangle inequality, as 
\begin{align*}
T(X)&= \Big(1-P_\mathrm{e}(X)-\frac1q\Big) + \sum_{x\ne\hat{x}} \left|P_{X}(x) - \frac1q\right|\\
&\ge \Big(1-P_\mathrm{e}(X)-\frac1q\Big) + \left|\sum_{x\ne\hat{x}} \left(P_{X}(x) - \frac1q\right)\right|\\
&= \Big(1-P_\mathrm{e}(X)-\frac1q\Big) + \frac{q-1}q - P_\mathrm{e}(X)\\
&= 2\left(\frac{q-1}q - P_\mathrm{e}(X)\right).
\end{align*}

\section{Proof of Lemma~\ref{lem:SP}}\label{sec:proofSP}
As before, it is again sufficient to prove an unconditional version of the inequalities in Lemma~\ref{lem:SP}.  
The unconditional version $S(X)$ of $S(X\mid Y)$ is defined as
\begin{align*}
S(X) &:= \frac1{q-1} \sum_{w\in\mathbb{F}_q^\times}\left|P^*_{X}(w)\right|
\end{align*}
where $P^*_X(w)$ denotes the unconditional version of $P^*_{X\mid Y}(w\mid y)$ defined as
\begin{align*}
P^*_X(w)&:= \sum_{z\in\mathbb{F}_q} P_{X}(z)\chi(wz).
\end{align*}

For the upper bound, one obtains
\begin{align*}
(q-1)S(X)&=
\sum_{w\in\mathbb{F}_q^\times} |P_{X}^*(w)|\le 
\sqrt{q-1}\sqrt{\sum_{w\in\mathbb{F}_q^\times} |P_{X}^*(w)|^2}\\
&= 
\sqrt{q(q-1)}\sqrt{\sum_{x\in\mathbb{F}_q} \left|P_{X}(x)-\frac1q\right|^2}.
\end{align*}
Here, the inequality is obtained from 
the Cauchy-Schwarz inequality $\|p_0^{q-1}\|_1 \le \sqrt{q} \|p_0^{q-1}\|_2$ 
which holds for $p_0^{q-1}\in\mathbb{C}^q$.
The last equality holds via Perseval's identity, 
i.e., since the Fourier transform is unitary up to the constant factor $\sqrt{q}$.
Let $t:=\lfloor 1/(1-P_\mathrm{e}(X))\rfloor$.
\begin{align}
&\sqrt{\sum_{x\in\mathbb{F}_q} \left|P_{X}(x)-\frac1q\right|^2}
\le
\Biggl(t\left|1-P_\mathrm{e}(X)-\frac1q\right|^2\nonumber\\
&\quad+\left|1-t(1-P_\mathrm{e}(X))-\frac1q\right|^2
+(q-t-1)\frac1{q^2}\Biggr)^{\frac12}\nonumber\\
&= \Biggl((1-P_\mathrm{e}(X))t\left((1-P_\mathrm{e}(X)) t - P_\mathrm{e}(X)\right)\nonumber\\
&\quad - t(1-P_\mathrm{e}(X)) + \frac{q-1}q\Biggr)^{\frac12}
\label{eq:L2}
\end{align}
Since~\eqref{eq:L2} is piecewise convex with respect to $P_\mathrm{e}(X)$, its concave hull is
\begin{align*}
&t(t+1)\Biggl[(t/(t+1)-P_\mathrm{e}(X))\sqrt{\frac{q-1}q - \frac{t-1}t}\\
&\quad + (P_\mathrm{e}(X)- (t-1)/t) \sqrt{\frac{q-1}q - \frac{t}{t+1}}\Biggr]
\end{align*}
for $P_\mathrm{e}(X)\in[0,(q-1)/q)$.
Since this is piecewise linear and convex, $t$ can be replaced by any $k=1,\dotsc,q-1$.
Note that the following smooth upper bound is obtained by replacing the first $(1-P_\mathrm{e}(X))t$ in~\eqref{eq:L2} by 1.
\begin{align*}
S(X) \le \sqrt{1 - \frac{q}{q-1}P_\mathrm{e}(X)}.
\end{align*}
The unconditional version of the lower bound in Lemma~\ref{lem:SP} 
is obtained via the triangle inequality, as 
\begin{align*}
(q-1)S(X)+1&=
\sum_{w\in\mathbb{F}_q} |P_{X}^*(w)|= 
\sum_{w\in\mathbb{F}_q} \left|\sum_{z\in\mathbb{F}_q} P_{X}(z) \chi(wz)\right|\\
&=\max_{a\in\mathbb{F}_q}\sum_{w\in\mathbb{F}_q} \left|\sum_{z\in\mathbb{F}_q} P_{X}(z) \chi(w(z-a))\right|\\
&\ge\max_{a\in\mathbb{F}_q}\left|\sum_{z\in\mathbb{F}_q} P_{X}(z)\sum_{w\in\mathbb{F}_q}  \chi(w (z-a))\right|\\
&=q\max_{a\in\mathbb{F}_q}P_{X}(a)
=q(1-P_\mathrm{e}(X)).
\end{align*}

\section{Proof of Lemma~\ref{lem:Sm}}\label{sec:proofS}
As in the argument for the binary case in~\cite[Chapter 5]{korada2009thesis}, 
MacWilliams identity is useful for the proof.
Let $H:= G^{-1}$ and $H_{\langle i\rangle} := [h_0,\dotsc,h_i]$ where $h_i$ is the $i$-th column of $H$.
Furthermore, we let the Fourier transform of the joint probability $P_{X, Y}$ 
be defined as $P_{X, Y}^*(w, y) := P_Y(y)P_{X\mid Y}^*(w\mid y)$.   
The generalized MacWilliams identity is obtained as follows. 
\begin{align*}
&P_{X^{(i)}, Y^{(i)}}(u_i, (u_0^{i-1}, y_0^{\ell-1}))\\
&= \sum_{x_{0}^{\ell-1}\in\mathbb{F}_q^\ell} 
\mathbb{I}\{x_0^{\ell-1}H_{\langle i\rangle}=u_0^{i}\}
\prod_{j=0}^{\ell-1}P_{X, Y}(x_j, y_j)
\\
&= \sum_{x_{0}^{\ell-1}\in\mathbb{F}_q^\ell} 
\prod_{j=0}^{i} \bigg[\frac1q\sum_{w_j\in\mathbb{F}_q}\chi\bigg(w_j \Big(\sum_{k=0}^{\ell-1} H_{kj} x_k-u_j\Big)\bigg)\bigg]\\
&\quad\cdot\prod_{j=0}^{\ell-1}\bigg[\frac1q\sum_{z_j\in\mathbb{F}_q} \chi(-z_jx_j) P^*_{X, Y}(z_j, y_j)\bigg]
\\
&= \frac1{q^{\ell+i+1}}\\
&\quad\cdot \sum_{z_{0}^{\ell-1}\in\mathbb{F}_q^\ell, w_0^{i}\in\mathbb{F}_q^{i+1}}
\prod_{j=0}^{\ell-1}\bigg[\sum_{x_j\in\mathbb{F}_q} \chi\bigg(x_j\Big(\sum_{k=0}^i H_{jk} w_k - z_j\Big)\bigg)\bigg]\\
&\quad\cdot\prod_{j=0}^{\ell-1}P^*_{X, Y}(z_j, y_j)
\prod_{j=0}^{i}\chi(-w_ju_j)
\\
&= \frac1{q^{i+1}} \sum_{z_{0}^{\ell-1}\in\mathbb{F}_q^\ell, w_0^{i}\in\mathbb{F}_q^{i+1}} 
\mathbb{I}\big\{w_0^iH_{\langle i\rangle}^{t}=z_0^{\ell-1}\big\}\\
&\quad\cdot\prod_{j=0}^{\ell-1}P^*_{X, Y}(z_j, y_j)
\prod_{j=0}^{i}\chi(-w_j u_j).
\end{align*}
Hence, the Fourier transform $P^{*}_{X^{(i)}, Y^{(i)}}$ of the joint probability 
$P_{X^{(i)}, Y^{(i)}}$ is given by 
\begin{align*}
&P^{*}_{X^{(i)}, Y^{(i)}}(w_i, (u_0^{i-1}, y_0^{\ell-1})) = \frac1{q^i}\\
&\quad\cdot\sum_{z_{0}^{\ell-1}\in\mathbb{F}_q^{\ell}, w_0^{i-1}\in\mathbb{F}_q^{i}}
\mathbb{I}\big\{w_0^{i-1}H_{\langle i-1\rangle}^{t}+w_i h_i^t=z_0^{\ell-1}\big\}\\
&\quad\cdot\prod_{j=0}^{\ell-1}P^*_{X, Y}(z_j, y_j)
\prod_{j=0}^{i-1}\chi(-w_j u_j).
\end{align*}
Then, one can derive the first inequality in Lemma~\ref{lem:Sm} as 
\begin{align}
&S_{\max}(X^{(i)}, Y^{(i)})\nonumber\\
&= \max_{w_i\in\mathbb{F}_q^\times}\sum_{y_0^{\ell-1}\in\mathcal{Y}^\ell, u_{0}^{i-1}\in\mathbb{F}_q^i}
|P^{*}_{X^{(i)}, Y^{(i)}}(w_i, (u_0^{i-1}, y_0^{\ell-1}))|\nonumber\\
&=
\max_{w_i\in\mathbb{F}_q^\times}\sum_{y_0^{\ell-1}\in\mathcal{Y}^\ell, u_{0}^{i-1}\in\mathbb{F}_q^i}
\Biggl|\frac1{q^i}\nonumber\\
&\quad\cdot\sum_{z_{0}^{\ell-1}\in\mathbb{F}_q^{\ell}, w_0^{i-1}\in\mathbb{F}_q^{i}}
\mathbb{I}\big\{w_0^{i-1}H_{\langle i-1\rangle}^{t}+w_ih_i^t=z_0^{\ell-1}\big\}\nonumber\\
&\quad\cdot\prod_{j=0}^{\ell-1}P^*_{X, Y}(z_j, y_j)
\prod_{j=0}^{i-1}\chi(-w_j u_j)
\Biggr|\nonumber\\
&\le
\max_{w_i\in\mathbb{F}_q^\times}\sum_{z_0^{\ell-1}\in\mathbb{F}_q^{\ell}, w_0^{i-1}\in\mathbb{F}_q^{i}} 
\mathbb{I}\{w_0^{i-1}H_{\langle i-1\rangle}^{t}+w_ih_i^t=z_0^{\ell-1}\}\nonumber\\
&\quad\cdot\prod_{j=0}^{\ell-1}\sum_{y\in\mathcal{Y}} \left|P^{*}_{X, Y}(z_j, y)\right|\nonumber\\
&\le q^{i} S_{\max}(X, Y)^{D_\mathrm{s}^{(i)}(G)}.
\label{eq:smax}
\end{align}
The last inequality in the above is obtained by observing that $z_0^{\ell-1}$ 
satisfying $w_0^{i-1}H_{\langle i-1\rangle}^{t}+w_ih_i^t=z_0^{\ell-1}$ 
should contain at least $D_\mathrm{s}^{(i)}(G)$ nonzero elements, 
and that $\sum_{y\in\mathcal{Y}}|P_{X, Y}^*(0, y)|=1$ holds.  

As for the second inequality in Lemma~\ref{lem:Sm}, one has 
\begin{align}
&S_{\min}(X^{(i)}, Y^{(i)})\nonumber\\
&= \min_{w_i\in\mathbb{F}_q^\times}\sum_{y_0^{\ell-1}\in\mathcal{Y}^\ell, u_{0}^{i-1}\in\mathbb{F}_q^i}
|P^{*}_{X^{(i)}, Y^{(i)}}(w_i, (u_0^{i-1}, y_0^{\ell-1}))|\nonumber\\
&=
\min_{w_i\in\mathbb{F}_q^\times}\sum_{y_0^{\ell-1}\in\mathcal{Y}^\ell, u_{0}^{i-1}\in\mathbb{F}_q^i}
\Biggl|\frac1{q^i}\nonumber\\
&\quad\cdot\sum_{z_{0}^{\ell-1}\in\mathbb{F}_q^\ell, w_0^{i-1}\in\mathbb{F}_q^i}
\mathbb{I}\{w_0^{i-1}H_{\langle i-1\rangle}^{t}+w_ih_i^t=z_0^{\ell-1}\}\nonumber\\
&\quad\cdot\prod_{j=0}^{\ell-1}P^*_{X, Y}(z_j, y_j)
\prod_{j=0}^{i-1}\chi(-w_j u_j)
\Biggr|\nonumber\\
&=
\min_{w_i\in\mathbb{F}_q^\times}
\max_{a_0^{i-1}\in\mathbb{F}_q^i}
\sum_{y_0^{\ell-1}\in\mathcal{Y}^\ell, u_{0}^{i-1}\in\mathbb{F}_q^i}
\Bigg|\frac1{q^i}\nonumber\\
&\quad\cdot\sum_{z_{0}^{\ell-1}\in\mathbb{F}_q^\ell, w_0^{i-1}\in\mathbb{F}_q^i}
\mathbb{I}\{w_0^{i-1}H_{\langle i-1\rangle}^{t}+w_ih_i^t=z_0^{\ell-1}\}\nonumber\\
&\quad\cdot\prod_{j=0}^{\ell-1}P^*_{X, Y}(z_j, y_j)
\prod_{j=0}^{i-1}\chi((a_j-w_j) u_j)
\Bigg|\nonumber\\
&\ge
\min_{w_i\in\mathbb{F}_q^\times}
\max_{a_0^{i-1}\in\mathbb{F}_q^i}
\sum_{y_0^{\ell-1}\in\mathcal{Y}^\ell}\nonumber\\
&\quad\Bigg|\sum_{z_{0}^{\ell-1}\in\mathbb{F}_q^\ell, w_0^{i-1}\in\mathbb{F}_q^i}
\mathbb{I}\{w_0^{i-1}H_{\langle i-1\rangle}^{t}+w_ih_i^t=z_0^{\ell-1}\}\nonumber\\
&\quad\cdot \prod_{j=0}^{\ell-1}P^*_{X, Y}(z_j, y_j)
\prod_{j=0}^{i-1}\bigg(\frac1q \sum_{u\in\mathbb{F}_q} \chi\big((a_j-w_j) u\big)\bigg)
\Bigg|\nonumber\\
&=
\min_{w_i\in\mathbb{F}_q^\times}
\max_{a_0^{i-1}\in\mathbb{F}_q^i}
\prod_{j=0}^{\ell-1}\sum_{y\in\mathcal{Y}}\left|P^*_{X, Y}((a_0^{i-1}H_{\langle i-1\rangle}^t+w_ih_i^t)_j, y)\right|\nonumber\\
&\ge
\min_{w_i\in\mathbb{F}_q^\times}
\max_{a_0^{i-1}\in\mathbb{F}_q^i}
\prod_{j=0}^{\ell-1}S_\mathrm{min}(X, Y)^{\mathbb{I}\{(a_0^{i-1}H_{\langle i-1\rangle}^t+w_ih_i^t)_j\not=0\}}\nonumber\\
&= S_{\min}(X, Y)^{D_\mathrm{s}^{(i)}(G)}
\label{eq:smin}
\end{align}
where the last equality in the above is obtained by noting that 
the maximization with respect to $a_0^{i-1}$ amounts to 
making the number of nonzero elements in 
$a_0^{i-1}H_{\langle i-1\rangle}^t+w_ih_i^t$ to be as small as possible.  

\section{Proof of Lemma~\ref{lem:ext}}\label{sec:proofext}
For the first equation, let us consider a $\sigma(B_1,\dotsc,B_n)$-measurable random process $\{\xi_n := \xi(\mathsf{X}_n,\mathsf{Y}_n)\}_{n=0,1,\dotsc}$ where
\begin{equation*}
\xi(X,Y):=\left\{(x,x')\in\mathbb{F}_q^2\mid Z_{x,x'}(X\mid Y)=0\right\}
\end{equation*}
and where
\begin{equation*}
Z_{x,x'}(X\mid Y) :=  \sum_{y\in\mathcal{Y}} \sqrt{P_{Y\mid X}(y\mid x)P_{Y\mid X}(y\mid x')}.
\end{equation*}
Then, $\{\xi_n\}_{n=0,1,\dotsc}$ is obviously a Markov chain.
The Markov chain $\{\xi_n\}_{n=0,1,\dotsc}$ has 
the empty set $\phi$ as the absorbing state, i.e.,
$\Pr(\xi_n = \phi\mid \xi_{n-1} = \phi)=1$.
Since any source accessible from the original source $(X,Y)$ by $G$ is also polarized by $G$,
$\phi$ is the unique accessible absorbing state.
Hence, $\lim_{n\to\infty}\Pr(\xi_n=\phi)=1$, proving the first equation of the lemma.  

The second equation is obtained in the same way.
Let us define 
\begin{equation*}
S_{w}(X\mid Y) :=  \sum_{y\in\mathcal{Y}} P_Y(y) \left|P^*_{X\mid Y}(w\mid y)\right|
\end{equation*}
and let $\{\eta_n := \eta(\mathsf{X}_n,\mathsf{Y}_n)\}_{n=0,1,\dotsc}$ be a $\sigma(B_1,\dotsc,B_n)$-measurable random process  where
\begin{equation*}
\eta(X,Y):=\left\{w\in\mathbb{F}_q\mid S_{w}(X\mid Y)=0\right\}.
\end{equation*}
Then, $\{\eta_n\}_{n=0,1,\dotsc}$ is a Markov chain since one obtains
from the derivations of~\eqref{eq:smax} and \eqref{eq:smin} in Appendix~\ref{sec:proofS} 
that
\begin{align*}
&\max_{z_0^{\ell-1}\in\mathcal{C}_i(w)} \prod_{j=0}^{\ell-1}S_{z_j}(X\mid Y)\le
S_{w}(X^{(i)}\mid Y^{(i)})\\
&\qquad\le
q^i\max_{z_0^{\ell-1}\in\mathcal{C}_i(w)} \prod_{j=0}^{\ell-1}S_{z_j}(X\mid Y)
\end{align*}
for any $w\in\mathbb{F}_q^\times$ and $i=0,\dotsc,\ell-1$
where $\mathcal{C}_i(w)$ is the affine space $\{\sum_{j=0}^{i-1} a_j h_j^t + w h_i^t\mid a_0^{i-1}\in\mathbb{F}_q^i\}$ defined on the basis of 
the columns of $G^{-1}:=[h_0, h_1, \dotsc, h_{\ell-1}]$.  
The superscript ${}^t$ here denotes transpose of a vector.  
Then, it holds that $\lim_{n\to\infty} \Pr(\eta_n = \phi)= 1$ 
due to the same reason as that for $\{\xi_n\}_{n=0,1,\dotsc}$, 
which proves the second equation of the lemma.

\bibliographystyle{IEEEtran}
\bibliography{IEEEabrv,ldpc}

\begin{IEEEbiographynophoto}{Ryuhei Mori}
received the B.E. degree from Tokyo Institute of Technology, Tokyo, Japan in 2008,
and the M.Inf.\ and D.Inf.\ degrees from Kyoto University, Kyoto, Japan in 2010 and 2013, respectively.
His research interests include information theory, computer science and statistical physics.
\end{IEEEbiographynophoto}

\begin{IEEEbiographynophoto}{Toshiyuki Tanaka}
received the B.E., M.E.,
and D.E.\ degrees in electronics engineering
from the University of Tokyo, Tokyo, Japan, in
1988, 1990, and 1993, respectively. From 1993
to 2005, he was with the Department of
Electronics and Information Engineering at
Tokyo Metropolitan University, Tokyo, Japan.
He is currently a professor at the Graduate
School of Informatics, Kyoto University, Kyoto,
Japan. He received DoCoMo Mobile Science
Prize in 2003, and Young Scientist Award from the Minister of
Education, Culture, Sports, Science and Technology, Japan, in 2005.
His research interests are in the areas of information and
communication theory, statistical mechanics of information
processing, machine learning, and neural networks.
He is a member of the IEEE, the Japanese Neural Network Society,
the Acoustical Society of Japan, the Physical Society of Japan,
and the Architectural Institute of Japan.
\end{IEEEbiographynophoto}

\end{document}